\documentclass[twocolumn,nofootinbib,amsmath,prd,aps,superscriptaddress,
tightenlines,preprintnumbers,nobalancelastpage]{revtex4}

\usepackage{slashed}
\usepackage{color, verbatim}
\usepackage{latexsym}
\usepackage{amsmath}
\usepackage{graphicx}

\usepackage{hyperref}

\usepackage{amssymb}
\usepackage{graphicx}
\usepackage{bm}
\usepackage{hyperref}
\usepackage{adjustbox}

\usepackage{graphicx}
\usepackage{latexsym}
\usepackage{amsmath}
\usepackage{mathrsfs}
\usepackage{hyperref}
\usepackage{enumerate}
\usepackage{adjustbox}
\usepackage{amssymb}
\usepackage{slashed}

\def\bea{\begin{eqnarray}}
\def\eea{\end{eqnarray}}

\begin{document}

\newcount\hour \newcount\minute
\hour=\time \divide \hour by 60
\minute=\time
\count99=\hour \multiply \count99 by -60 \advance \minute by \count99
\newcommand{\mydate}{\ \today \ - \number\hour :00}

\preprint{IPPP/18/104}
\title{\Large Gravitational wave and collider probes of\\[0.1cm] a triplet 
Higgs 
sector 
with a low cutoff}

\author{Mikael Chala$^a$, Maria Ramos$^b$ and Michael Spannowsky$^a$\\\vspace{0.4cm}
\it {$^a$Institute of Particle Physics Phenomenology, Physics Department, Durham 
University, Durham DH1 3LE, UK}\\[0.2cm]
\it {$^b$Laborat\'orio de Instrumenta\c{c}\~ao e F\'isica Experimental de Part\'iculas, Departamento de F\'isica
da Universidade do Minho, Campus de Gualtar, 4710-057 Braga, Portugal}
}

\begin{abstract}
We study the scalar triplet extension of the standard model with a low cutoff, 
preventing large corrections to the quadratic masses that would otherwise 
worsen the hierarchy problem. We explore the reach of LISA to test the 
parameter space region of the scalar potential (not yet excluded by Higgs 
to diphoton measurements) in which the electroweak phase transition is strongly 
first-order and produces sizeable gravitational waves. We also demonstrate that 
the collider phenomenology of the model is drastically different from its 
renormalizable counterpart. We study the reach 
of the LHC in ongoing searches and project bounds for the HL-LHC. Likewise, 
we develop a dedicated analysis to test the key but still unexplored 
signature of pair-production of charged scalars decaying to 
third-generation quarks: $pp\rightarrow t\overline{b} (\overline{t}b), 
b\overline{b}$. These 
results apply straightforwardly to other extensions of the Higgs sector such as 
the 2HDM/MSSM.
\end{abstract}

\maketitle

\tableofcontents

\section{Introduction}
Hyperchargeless triplet scalars $\Phi$ arise in a variety of models
of new physics. They include 

\textit{(i)} theories of grand
unification (GUT), where scalar multiplets, often transforming in the
adjoint representation of the GUT group, break spontaneously the GUT symmetry. A simple example is the
$\mathbf{24}$ in $SU(5)$, which decomposes as $(1,1)_0+(1,\mathbf{3})_0+\cdots$
under the Standard Model (SM) gauge group $SU(3)_c\times SU(2)_L\times U(1)_Y$, 
therefore delivering a scalar triplet.
Likewise, the $\mathbf{45}$ representation of $SO(10)$
contains the $\mathbf{24}$ of  $SU(5)$ and therefore a SM triplet as well.

\textit{(ii)} Supersymmetric (SUSY) models. As a matter of fact, the triplet
extension of the MSSM is one of the simplest options to alleviate
the little hierarchy problem~\cite{Espinosa:1991wt,DiChiara:2008rg}. 

\textit{(iii)} Composite
Higgs models (CHM). The scalar sector of most CHMs is non
minimal. It includes a hyperchargeless triplet in one of the
two
simplest cosets admitting an UV completion \textit{\`a la QCD}
in four dimensions, \textit{viz.} 
$SU(5)/SO(5)$~\cite{Vecchi:2013bja,Ferretti:2016upr}.
Moreover, models based on $SO(7)/G_2$~\cite{Chala:2012af} provide exactly one
triplet in addition to the Higgs boson.

Therefore, the phenomenology of such triplet is not dictated
by the renormalizable Lagrangian. The latter has to be
instead supplemented with effective operators encoding the
effects of the heavier resonances (SUSY partners, composite
states, etc.),
which can modify drastically
the dynamics of $\Phi$. To demonstrate this, we
will work under the assumption that the triplet does not get a (custodial 
symmetry breaking)
vacuum expectation value (VEV). This limit can be naturally enforced assuming 
the
triplet
is a CP-odd scalar and CP is conserved in the Higgs sector.
At the renormalizable level, the Lagrangian becomes accidentally
$\mathbb{Z}_2$ symmetric, \textit{i.e.} $\Phi\rightarrow -\Phi$, making the neutral component of
the triplet a potential dark matter candidate. The charged
components are in turn long-lived. The corresponding
phenomenology has been studied in 
Refs.~\cite{Cirelli:2005uq,FileviezPerez:2008bj,Carmona:2015haa}. 
However,
the effective operators make all components decay promptly
even if the cutoff is $f\sim$ several TeV at which new
resonances are out of the reach of current facilities.
A much larger cutoff would introduce too large corrections
also to the triplet mass, worsening the hierarchy problem.
\footnote{(The fine-tuning scales roughly as 
$\Delta \sim m_h^2/f^2$, with $m_h$ the Higgs mass~\cite{Panico:2015jxa}. $f=1$ 
TeV gives already $\Delta\sim 1\,\%$, and this falls below the permile level 
for $f> 4$ TeV.)}
In this article, we study probes of current and future
colliders to this more natural version of the inert triplet
model (ITM).

The extended Higgs sector modifies also the electroweak (EW) phase 
transition (EWPT). Thus, we extend previous studies in this 
respect~\cite{Blinov:2015sna,Inoue:2015pza} 
computing the reach 
of future detectors
for gravitational waves that originate in the
production, evolution and eventual collisions of bubbles of
vacuum in a first-order phase transition. The paper is organized as follows.
We introduce the model in section~\ref{sec:model}. We discuss the dynamics of 
the EWPT in section~\ref{sec:ewpt}. We explore collider signatures in 
section~\ref{sec:collider}. In section~\ref{sec:analysis} we propose an LHC 
analysis that has not been yet worked out experimentally for probing the 
key 
channel $pp\rightarrow t\overline{b} (\overline{t}b), b\overline{b}$. We study 
signal and background and provide 
prospects for the HL-LHC, namely the LHC running at a center of mass energy 
(c.m.e) $\sqrt{s} = 13$ TeV with 
integrated luminosity $\mathcal{L} = 3$ ab$^{-1}$. We conclude in 
section~\ref{sec:conclusions}.

\section{Model}\label{sec:model}
The Lagrangian of the CP-odd scalar triplet when it is assumed embedded in 
an UV theory, takes the form:
\begin{align}\nonumber\label{eq:lag}
L = &\frac{1}{2}|D_\mu\Phi|^2 
-\bigg\lbrace\frac{1}{2}\mu_\Phi^2|\Phi|^2 + \frac{1}{2}\lambda_{H\Phi}|H|^2 
|\Phi|^2 + \frac{1}{4}\lambda_\Phi|\Phi|^4\bigg\rbrace
\\
& + 
\frac{1}{f}\bigg\lbrace 
\text{i}\,\overline{q_L^i}\bigg[c^u_{ij}(\tilde{H}\Phi) u^j_R  + 
c^d_{ij}(H\Phi)d_R^j\bigg] + \text{h.c.}\bigg\rbrace~,
\end{align}
where $H=(h^+, h_0 = (h+v)/\sqrt{2})$ and $\Phi = (\phi^+, -\phi^0, \phi^-)$. 
We will assume flavour diagonal couplings: $c^{u(d)}_{ij}\sim 
c y_{u(d)}\delta_{ij}$ with $c$ a constant and $y$ Yukawa. 
We will comment on departures from this assumption in the 
conclusions.
The relevant parameter is therefore the ratio $c/f$.
The product $\tilde{H}\Phi$ $(H\Phi)$ stands for the doublet in the 
$SU(2)_L\times U(1)_Y$
decomposition $\mathbf{2}_{-(+)1/2}\times\mathbf{3}_0 = \mathbf{2}_{-(+)1/2} + 
\mathbf{4}_{-(+)1/2}$. Explicitly:
\begin{equation}
 \tilde{H}\Phi = \bigg(\begin{array}{c}
                              \phi_0 h_0^* - \sqrt{2}\phi^+ h^- \\
                              \phi_0 h^- - \sqrt{2} \phi^- h_0^*
                             \end{array}\bigg)~,
\end{equation}%
and analogously for $H\Phi$ upon the replacement $h_0^*\rightarrow h^+$ and 
$h^- \rightarrow -h_0$. 	
Therefore, in the unitary gauge after EWSB, we obtain:
\begin{align}
 L \supset  \dfrac{v}{\sqrt{2}}\frac{c}{f}&\bigg\lbrace \text{i}y^t\phi_0 
\overline{t}\gamma_5 t - \text{i}y^b\phi_0 \overline{b}\gamma_5 b \\
 &\bigg[-\sqrt{2} \text{i} \phi^- \overline{b}(y^t P_R + y^b P_L )t + 
\text{h.c.}\bigg]\bigg\rbrace~,
\end{align}
with $v\sim 246$ GeV 
and the sum extends to the first and second families of quarks, that we will 
denote collectively by $q$. 
Couplings to the leptons could be also present. We neglect them in this 
analysis.
\begin{figure}[t]
 \includegraphics[width=\columnwidth]{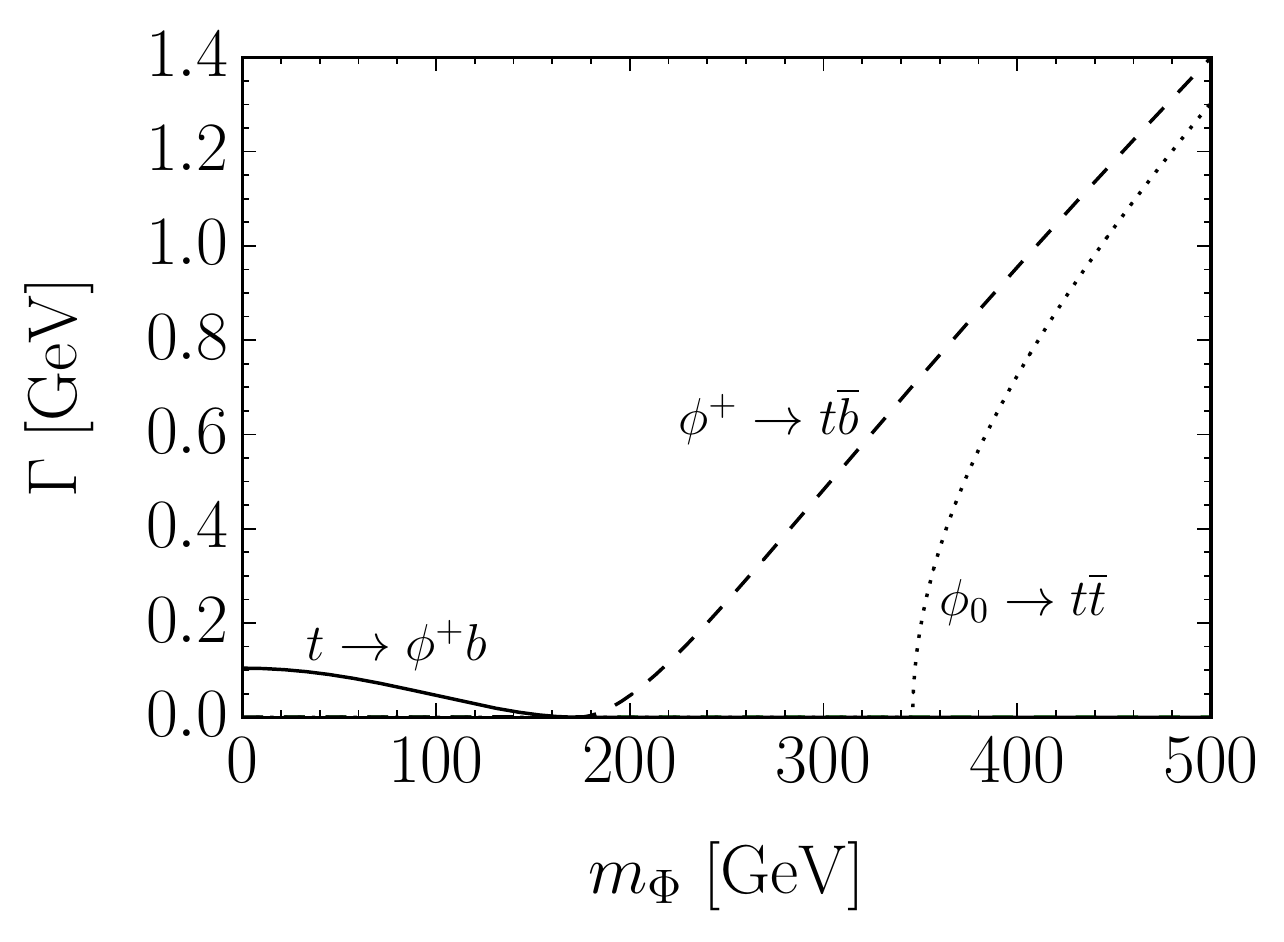}
 \caption{\it Exotic decay widths of the different heavy 
particles in our model for $f = 1$ TeV.}\label{fig:width1}
\end{figure}

$\phi^{0(\pm)}$ can decay 
into SM quarks. Likewise, for  $m_t> m_\Phi$, the top quark can decay into the 
triplet and a bottom quark. ($m_t$ stands for the top quark mass, whereas 
$m_\Phi$ is the physical mass of $\Phi$; at tree level 
$m_\Phi^2 = \mu_\Phi^2 + \lambda_{H\Phi}v^2/2$.) The following relations hold: 
\begin{align}\label{eq:widths}
 \Gamma \left(\phi_0 \rightarrow q \overline{q}\right) &= \frac{3 y_q^2 
v^2}{16\pi}\frac{c^2}{f^2} m_\phi \sqrt{ 1 - \frac{4 m_q^2}{m_\phi^2}}~,\\
  \Gamma \left(\phi^+ \rightarrow q \overline{q^\prime} \right)  &= \frac{3 
(y_q^2 + y_{q^\prime}^2) v^2}{16 \pi}\frac{c^2}{ f^2}  m_\phi  
\left[1-\frac{m_q^2}{m_\phi^2}\right]^2~,\\
 \Gamma \left(t \rightarrow \phi^+ b\right)  &= \frac{(y_b^2 + y_t^2) v^2}{32 
\pi}\frac{c^2}{ f^2}  m_t \left[1 - \frac{m_\phi^2}{m_t^2}\right]^2~.
\end{align}
In the second equation we are assuming $m_q \gg m_{q^\prime}$, which is 
normally the case if the former is an up quark and the later a down quark. Note 
that decays into the light quarks are dominant if channels involving 
the top quark are kinematically closed. Partial widths as a function of 
$m_\Phi$ are depicted in Fig.~\ref{fig:width1}.

Note also that, had we assumed a CP violating trilinear term in the 
potential, $\sim 
\kappa\Phi H^2$, this term would induce a VEV for the triplet, $v^\prime\sim 
\kappa v^2/m_\Phi^2$. The triplet could also decay into the Higgs degrees of 
freedom, the corresponding width scaling as $\Gamma\sim (v^\prime/v^2)^2 
 m_\Phi^3$. 
Given that $v^\prime$ modifies the $\rho$ parameter, it is bounded to be 
$v^\prime \lesssim $ GeV~\cite{Tanabashi:2018oca}. 
Therefore, the 
corresponding decay 
would still be subdominant with respect to those suppressed by $v/f$ even for 
$f\sim 10$ TeV.

This setup can be easily accommodated in a SUSY framework. The minimal model
consists of the MSSM extended with a supermultiplet $\Sigma$ with quantum 
numbers $(1, \mathbf{3})_0$; see Ref.~\cite{Delgado:2013zfa}. The most 
general and renormalizable superpotential is the MSSM superpotential extended by
\begin{equation}
 \Delta W = \mu H_1 H_2 + \lambda H_1 \Sigma H_2 + \frac{1}{2}\mu_\Sigma \text{tr} \Sigma^2~,
\end{equation}
with $H_1$ and $H_2$ the two doublet superfields. Likewise, the 
extra soft-breaking Lagrangian reads
\begin{equation}
 L_{\text{SB}} = m_4^2 \text{tr} \Sigma^\dagger\Sigma + \bigg[B_\Sigma\text{tr}\Sigma^2 + \lambda A_\lambda H_1\Sigma H_2 + \text{h.c.}\bigg]~.
\end{equation}
In the limit $\lambda\rightarrow 0$, $\lambda\mu_\Sigma\rightarrow 
\text{finite}$, the fermionic partner in $\Sigma$ decouples and the model is 
approximately inert with 
\begin{align}
 \mu_\Phi^2 & \sim m_4^2 + \mu_\Sigma^2 + B_\Sigma\mu_\Sigma~,\\
 \kappa_{\Phi} &\sim -\sqrt{2}\lambda\mu_\Sigma~.
\end{align}
In the previous expression, $\kappa_{\Phi}$ stands for the trilinear coupling
in $\kappa_\Phi \Phi H H_{2}$. (With a slight abuse of notation, we are 
denoting here by $H_2$ the scalar component of this superfield.)
After integrating $H_2$ out, we get
\begin{align}
 \lambda_{H\Phi} \sim \dfrac{(\lambda\mu_{\Sigma})^2}{m_2^2}~, \quad 
\frac{c^{u(d)}}{f}\sim 
y_2^{u(d)}\frac{\lambda\mu_\Sigma}{m_2^2}~,
\end{align}
with $m_2$ the mass of $H_2$ and $y_2^{u(d)}$ its Yukawa couplings. 

The Lagrangian above can also arise naturally in CHMs, in which 
both $H$ and $\Phi$ are pNGBs originated in the (approximate) symmetry breaking 
pattern $\mathcal{G}\rightarrow \mathcal{H}$ at a scale $f$ at which a new 
strong sector confines. The smallest realization of this setup relies on 
$SO(7)\rightarrow G_2$. The global symmetry is only approximate because 
it is explicitly broken by loops of SM gauge bosons, as well as by linear 
mixings between the left- and right-handed top quark fields and 
composite operators $\mathcal{O}_{L,R}$. The latter transform in 
representations of $SO(7)$. If $\mathcal{O}_L\sim \mathbf{35}$ and 
$\mathcal{O}_R\sim 1$, one obtains the Lagrangian~\cite{Ballesteros:2017xeg}
\begin{equation}
\label{Lchm}
 L \sim y_u\overline{q_L} \tilde{H}  \bigg[1 + \frac{\gamma}{f}\Phi + 
\mathcal{O}(1/f^2)\bigg]u_R~,
\end{equation}
and similarly for other quarks. $\gamma$ parametrises the degree of 
mixing of $q_L$ with the two doublets in the $\mathbf{35}$. (Under $G_2$, 
$\mathbf{35} = 1+\mathbf{7}+\mathbf{27}$, whereas $\mathbf{7} 
=(\mathbf{2},\mathbf{2})+(\mathbf{3},1)$ and $\mathbf{27} 
=(1,1)+(\mathbf{2},\mathbf{2})+(\mathbf{3},\mathbf{3})+(\mathbf{4},\mathbf{2}
)+(\mathbf{5},1)$ under the custodial symmetry group $SU(2)_L\times 
SU(2)_R$.) 

\begin{figure}[t]
 \begin{center}
  \includegraphics[width=0.32\columnwidth]{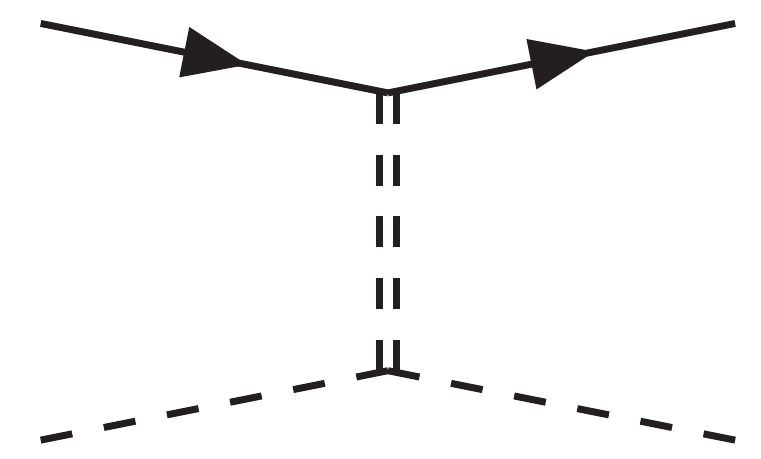}
  \includegraphics[width=0.32\columnwidth]{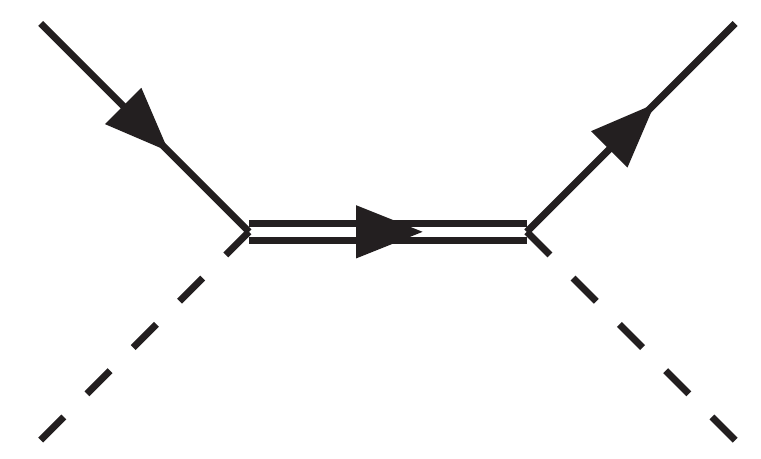}
  \includegraphics[width=0.32\columnwidth]{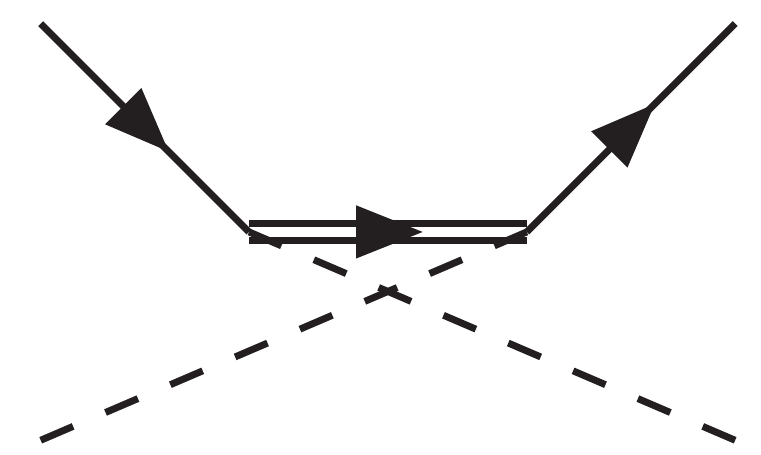}
 \end{center}
\caption{\it Generation of the effective operator in Eq.~\ref{eq:lag} after 
integrating out a heavy scalar with with quantum numbers $(1, 
\mathbf{2})_{1/2}$ (left), a heavy
vector-like quark  
$\sim(\mathbf{3},\mathbf{3})_{{-2/3}\, (1/3)}$ (center) or 
$\sim(\mathbf{3},\mathbf{2})_{-1/6}$ (right) at a mass scale $M\sim 
f$. From left to right and top to bottom, the 
different fields are: $q_L$, $t_R$, $H$, $\Phi$.}~\label{fig:EFTs}
\end{figure}

The elementary ITM can also get $1/f$ corrections 
provided it is extended with new vector-like quarks with quantum numbers 
$(\mathbf{3},\mathbf{3})_{-2/3\,(1/3)}$, $(\mathbf{3},\mathbf{2})_{-1/6}$; 
and/or 
with scalar doublets with quantum numbers $(1,\mathbf{2})_{1/2}$. The effective 
operator in Eq.~\ref{eq:lag} is then generated after integrating the heavy modes 
at the mass scale $M\sim f$; see Fig.~\ref{fig:EFTs}. This list exhausts the 
possible tree-level weakly-coupled UV completions of the ITM that fit into our 
phenomenological framework.

LEP operated at $\sqrt{s} = 209$ GeV, excluding $\Phi$ masses below $\sim 
100$ GeV. (This limit is however slightly model 
dependent; other 
sources of new physics could weaken it to even $\sim 75$ 
GeV~\cite{Egana-Ugrinovic:2018roi}.) We will therefore restrict our analysis to the mass 
range $100 
< m_\Phi < 500$ GeV.

\section{The electroweak phase transition}\label{sec:ewpt}
The new scalar potential modifies the EWPT, which in the SM is a cross 
over. In the region of the parameter space where it is first order and 
strong, gravitational waves can be produced via nucleation and eventual 
collision of bubbles of symmetry-breaking vacuum. In order to explore this 
phenomenology, we study the evolution 
of the one-loop effective potential at finite temperature:
\begin{equation}
 V = V_{\text{tree}} + \Delta V_\text{CW} + \Delta V_T~ + C.
\end{equation}
$C$ is a constant fixed so that $V$ vanishes at the origin of the field space. 
$V_{\text{tree}}$ stands for the tree-level potential. $V_\text{CW}$ is the 
one-loop correction at zero temperature in $\overline{\text{MS}}$ and Landau 
gauge, namely
\begin{align}
\Delta V_\text{CW} &= \frac{1}{64 \pi^2} \sum_i (\pm) n_i m_i^4 
\bigg[ \log{\frac{m_i^2}{v^2}} - c_i \bigg]~,
\end{align}
where $i$ runs over all bosons ($+$) and fermions $(-)$. The factor $n_i$ 
denotes de number of degrees of freedom of the field $i$, while $c_i$ is $5/6$ 
for gauge bosons and $3/2$ otherwise. The field-dependent masses squared of the 
spectator fields are
\begin{align}
 m_W^2 &= \frac{1}{4} g^2 (h^2 + 4 \phi_0^2)~, \\
 m_Z^2 &= \frac{1}{4} (g^2 + g'^2) h^2~, \\
 m_{G^{0, \pm}}^2 &= -\mu_H^2 + \lambda_H h^2 + \frac{1}{2}\lambda_{H\Phi} 
\phi_0^2~,\\
m_{\phi^\pm}^2 &= 
\mu_\Phi^2+\frac{1}{2}\lambda_{H\Phi}h^2+\lambda_\Phi\phi_0^2~,\\
m_t^2 &=  \frac{1}{2} y_t^2 h^2~.
\end{align}
In addition, we have two more field dependent masses squared, $m_1^2$ and 
$m_2^2$, given by the eigenvalues of the mixing matrix
\begin{align}
  &\mathcal{M}^2=\\\nonumber
  &\bigg[\begin{array}{cc}
        -\mu_H^2 + 3 \lambda_H h^2 + \frac{1}{2} \lambda_{H\Phi} \phi_0^2~ & 
\lambda_{H\Phi} h \phi_0\\
        \lambda_{H\Phi} h \phi_0 & \mu_\Phi^2 + 3 \lambda_\Phi \phi_0^2 + 
\frac{1}{2} \lambda_{H\Phi}  h^2
       \end{array}\bigg]
\end{align}
Finally, the finite temperature corrections read
\begin{align}
 \Delta V_{T} =& \frac{T^4}{2\pi^2}\sum_i (\pm)n_i \int_0^\infty y^2 \log{\bigg[1\mp e^{-\sqrt{\frac{m_i^2}{T^2}+y^2}}\bigg]}~.
\end{align}
As input parameters, we take $m_\Phi$, $\lambda_{H\Phi}$ and $\lambda_\Phi$. 
The remaining three parameters in the tree level potential are numerically 
obtained after requiring $V_\text{tree} + \Delta V_\text{CW}$ to have a 
extreme at $\langle h\rangle = v, \langle \phi_0\rangle = 0$, at which the 
physical 
Higgs and $\Phi$ masses are $m_h \sim 125$ GeV and $m_\Phi$, respectively. In 
other words:
\begin{align}
 \frac{\partial V}{\partial h} = 0~, \quad \frac{\partial^2 V}{\partial 
h^2} = m_h^2~, \quad \frac{\partial^2 V}{\partial \phi_0^2} = m_\Phi^2~.
\end{align}
At tree level, $(v, 0)$ is guaranteed to be an extreme provided 
$\lambda_{\Phi}, \lambda_{H\Phi}>0$ and $\mu_\Phi^2 > -1/2 v^2 
\lambda_{H\Phi}$.
A comparison between tree and loop level values of $\mu_H$, $\mu_\Phi$ and 
$\lambda_H$ in a set of benchmark inputs can be seen in 
Tab.~\ref{tab:comparison}. 

\begin{figure}[t]
 \includegraphics[width=\columnwidth]{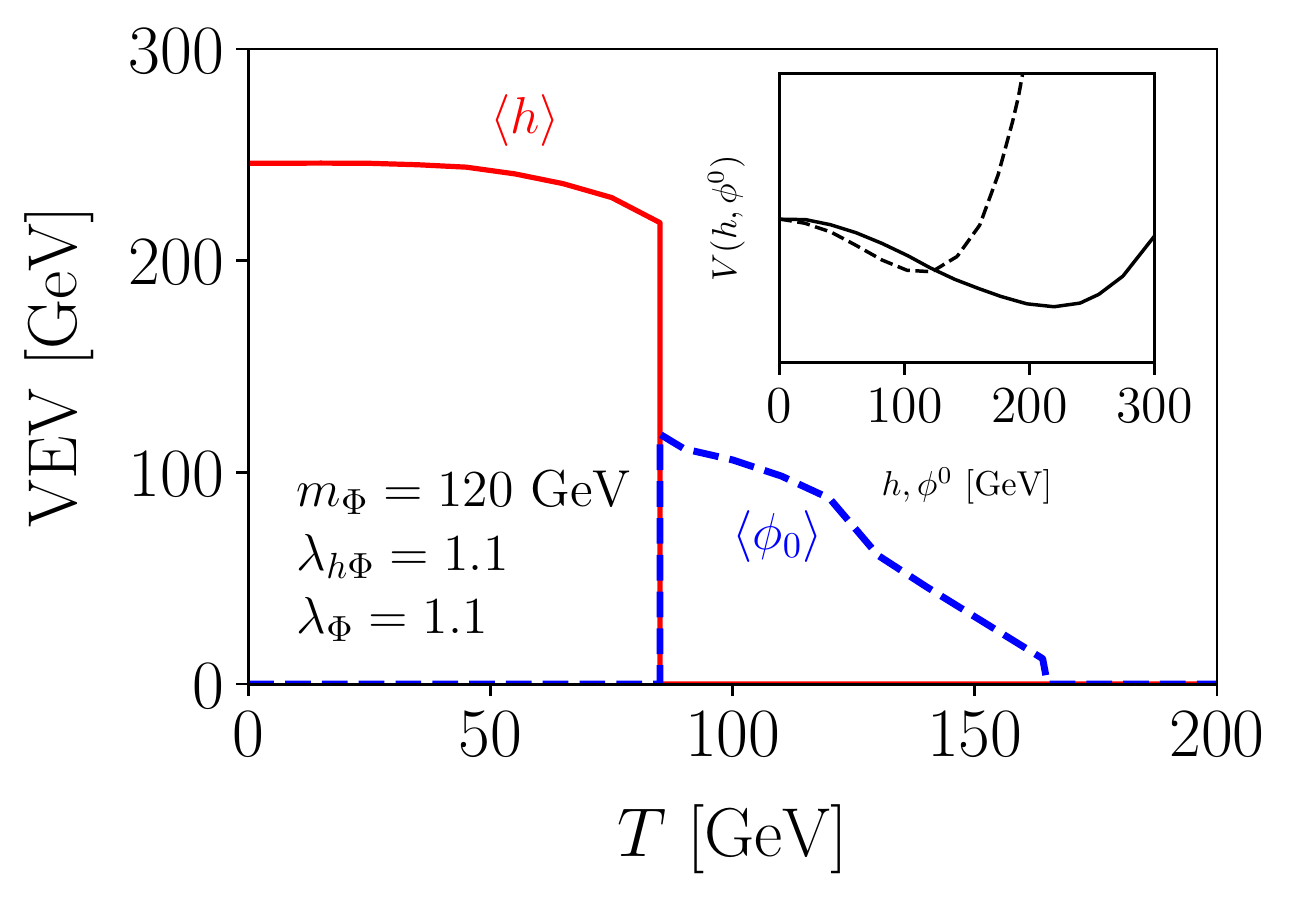}
 \caption{\it Evolution of the VEV with the temperature for the parameter
 space point $m_\Phi = 120$ GeV, $\lambda_{h\Phi}=1.1$, $\lambda_\Phi=1.1$. At 
high temperatures, the EW symmetry is restored. It is
 spontaneously broken to $(\langle h\rangle,\langle\phi_0\rangle) \sim (0, 10)$ 
GeV at $T\sim 160$ GeV, evolving until $T_n\sim 85$ GeV at which the step $(0, 
120)\rightarrow (220,0)$ GeV takes place.
 This latter transition is clearly strong; $\langle h\rangle/T_n > 1$. The 
 shape of the potential at $T_n$ is also shown.}\label{fig:evol}
\end{figure}
\begin{figure}[t]
 \includegraphics[width=\columnwidth]{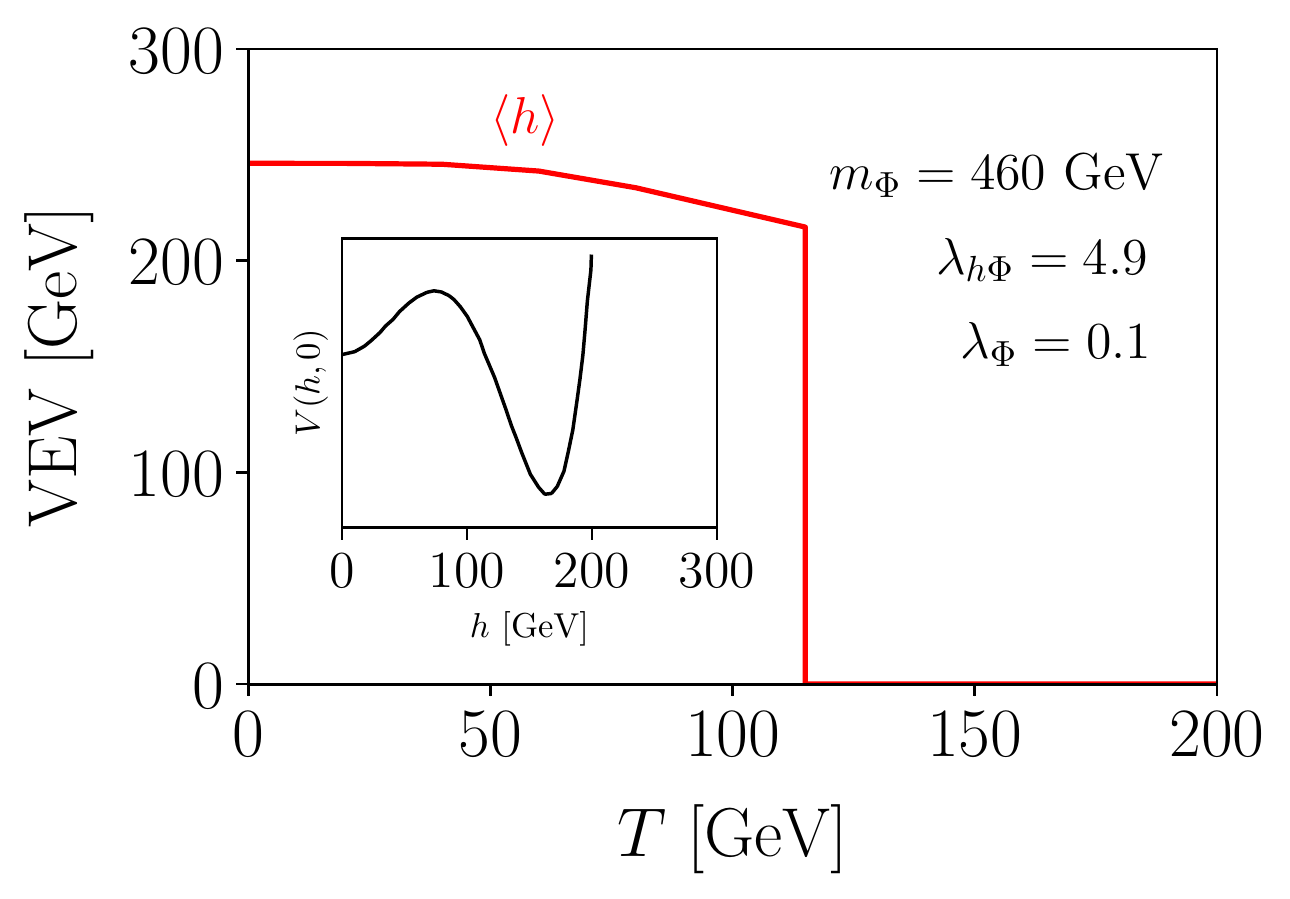}
 \caption{\it Same as Fig.~\ref{fig:evol} but for the 
parameter space point $m_\Phi=460$ GeV, $\lambda_{H\Phi}=4.9$, 
$\lambda_\Phi=0.1$. The EWPT proceeds in one step at $T_n = 115$ GeV in this 
case. $\langle\phi_0\rangle$ vanishes for all values of $T$.}\label{fig:evol1}
\end{figure}

At high temperatures, the EW symmetry is restored. For certain values 
of 
the model parameters, the transition between $\langle h\rangle = 
0\rightarrow\langle h \rangle = v_n$ is not smooth as in the SM, but rather 
first order. One example is given in Fig.~\ref{fig:evol}. In this 
case, the EWPT proceeds in two steps. An example of a first-order EWPT in one 
step is shown in Fig.~\ref{fig:evol1}. We will 
denote by $T_n$ the nucleation temperature, namely the temperature at which the 
Higgs 
first order phase transition takes place. 
This is determined by the condition $S_3/T_n \sim 100$, where $S_3$ stands 
for the action of
the thermal transition between vacua~\cite{Quiros:1999jp,Laine:2016hma}.

\begin{figure}
 \includegraphics[width=\columnwidth]{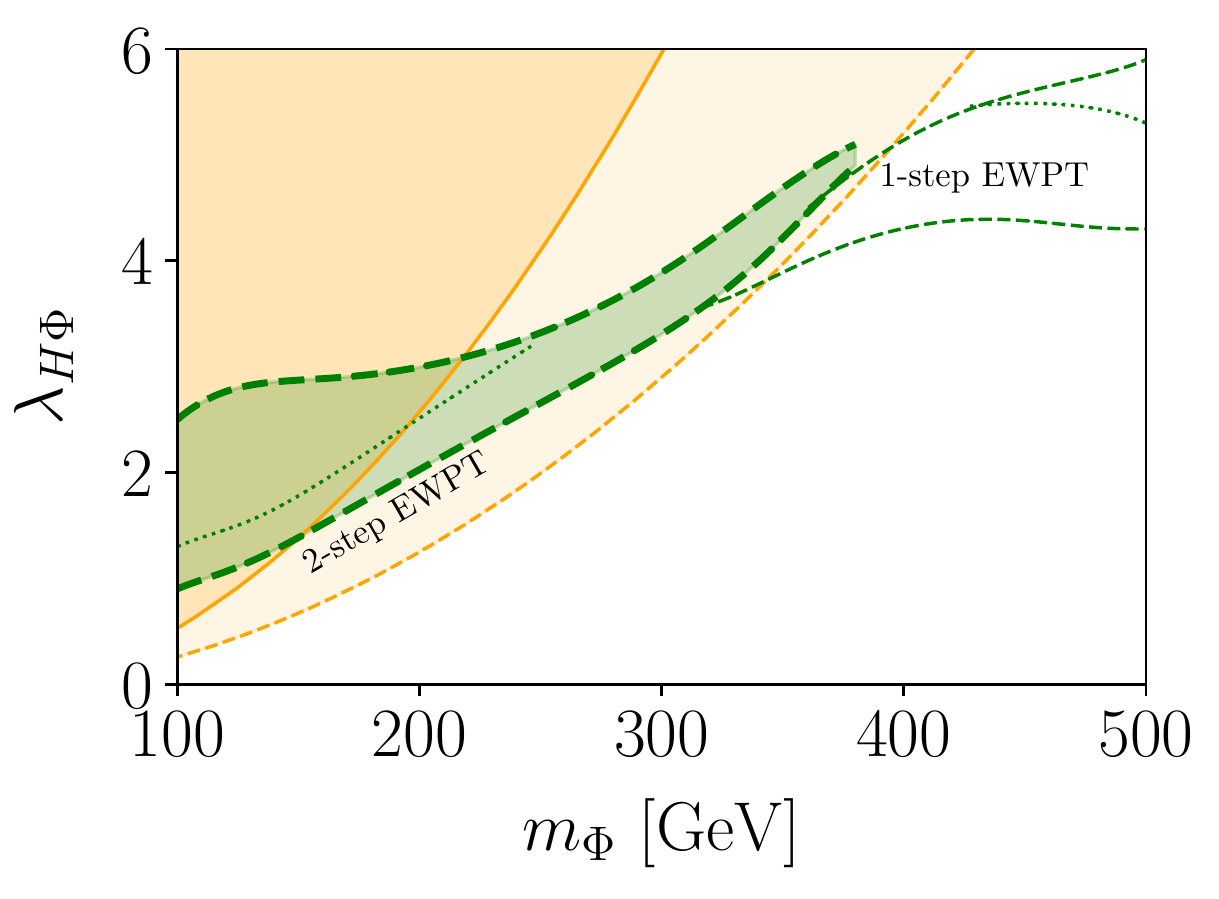}
 \caption{\it The region in the plane $(m_\Phi, \lambda_{h\Phi})$ 
where the EWPT is strongly first order is enclosed by the two 
dashed green lines, each one labeling the structure of the phase transition (1- or 2-step).
The sub-regions above the dotted lines can be tested at LISA.
The area enclosed by the solid (dashed) orange line is (could be) excluded at 
the $95$ \% CL by 
current (future) measurements of the Higgs to photon width. }\label{fig:ewpt}
\end{figure}

In the 
region enclosed by the dashed green line  in the 
plane ($\lambda_{H\Phi}, m_\Phi)$ of Fig.~\ref{fig:ewpt}, $v_n/T_n > 1$; 
\textit{i.e.} the 
phase 
transition is said to be \textit{strong}.  The nature 
of the strongest phase transition (one or two steps) is also labelled. The way 
we performed the scan 
is as follows: We varied $m_\Phi$ in the range $[100, 500]$ GeV in steps of 
$20$ GeV. We varied $\lambda_{H\Phi}$ in the range $[0.1, 10]$ in 
steps of $0.1$. For each pair $(m_\Phi, \lambda_{H\Phi})$, we found the value 
of 
$\lambda_\Phi$ in $[0.1, 0.3, \cdots, 10]$ maximizing $v_n/T_n$. The points 
with smallest value of $\lambda_{H\Phi}$ are interpolated using straight lines. 
Likewise for those with largest value of this 
coupling. The resulting 
lines are further smoothed according to the \textit{bezier} method using 
\texttt{Gnuplot}.

Note that at $T > 0$, the triplet squared term reads $\sim 
\mu^2_\Phi + T^2$, which can not be negative for any value of $\mu^2_\Phi > 0$. 
Therefore, the 2--step EWPT can only occur if the triplet minimum is present at 
$T=0$. Moreover, for a fixed $m_\Phi$, there is a minimum $\lambda_{H\Phi}$ 
below which $\mu^2_\Phi$ is not negative. Likewise, there is a maximum value of 
the coupling above which the potential at the triplet minimum, $V(0, \langle 
\phi_0\rangle)\sim -|\mu_\Phi|^4/\lambda_\Phi$, is deeper than the Higgs one, 
the theory being therefore unstable. Altogether, they explain the bounded shape 
of the figure above.

It is also 
well known that strong first order phase transition, resulting from 
non-standard Higgs sectors, produce gravitational 
waves~\cite{Choi:1993cv,Ashoorioon:2009nf,Enqvist:2014zqa,Caprini:2015zlo,
Kakizaki:2015wua,Schwaller:2015tja,Jaeckel:2016jlh,Huang:2016odd,Hashino:2016rvx,Tenkanen:2016idg,Huang:2016cjm,
Artymowski:2016tme,Huang:2017laj,Hashino:2016xoj, Vaskonen:2016yiu,Chao:2017vrq,
Hindmarsh:2017gnf,Kobakhidze:2016mch,Beniwal:2017eik,
Kobakhidze:2017mru,Cai:2017tmh,Bian:2017wfv,Huang:2017kzu,Croon:2018erz,Mazumdar:2018dfl,Hashino:2018wee,
Ahriche:2018rao,Beniwal:2018hyi,Huang:2017rzf,Breitbach:2018ddu}. They are roughly 
characterized by the normalized 
latent heat of 
the 
phase transition
\begin{equation}
 \alpha \sim \frac{\epsilon(T_n)}{35 T_n^4}~;
\end{equation}
(with $\epsilon(T_n)$ the latent
heat at $T_n$), 
and by the inverse duration time of the phase transition,
\begin{equation}
 \frac{\beta}{H} \sim T_n \frac{d}{dT}\frac{S_3}{T}\sim 
T_n\frac{\Delta(S_3/T)}{\Delta T}~.
\end{equation}
We computed these quantities using 
\texttt{CosmoTransitions}~\cite{Wainwright:2011kj}. $\Delta(S_3/T)$ is 
estimated finding the two values of $T$ for which $S_3/T = 100, 200$ GeV, 
respectively. (Therefore, $\Delta(S/T) = 100$ GeV.) We warn that, due to the 
rapid growth of $S_3/T$ with $T$, the linear estimation of the derivative 
can be sensibly overestimated. Given that small values of $\beta/H$ give 
rise to stronger gravitational waves, our results are conservative. The region 
of the parameter 
space that we estimate it can 
be tested by the future gravitational wave observatory LISA lies above the 
dotted green line in Fig.~\ref{fig:ewpt}. The points in this are lead 
to $\alpha, \beta$ within the region ``C1'' of Ref.~\cite{Caprini:2015zlo} for 
$T_n = 100$ GeV. (The bubble velocity is close to unity in good 
approximation. Also, we have neglected the effect that sounds waves might 
be not ``long-lasting'', what could weaken the gravitational wave 
signal~\cite{Ellis:2018mja}.)

\begin{table}[t]
\begin{center}
 \begin{tabular}{ccc|cccccc}
 \hline\hline
 $m_\Phi$ & $\lambda_{H\Phi}$ & $\lambda_\Phi$ & $(\mu_H^2)^0$ & 
$(\mu_\Phi^2)^0$ & 
$\lambda_H^0$ & $\mu_H^2$ & $\mu_\Phi^2$ & $\lambda_H$ \\\hline
120 & 1.1 & 1.1 & 7812.5 & -18883.8at & 0.13 & 9050.9 & $-16600.6$ & 0.127\\
200 & 2.0 & 1.0 & 7812.5 & -20516.0 & 0.13 & 8284.1 & $-17069.8$ & 0.125\\
320 & 3.5 & 1.5 & 7812.5 & -3503.0 & 0.13 & 5443.6 & 129.3 & 0.086\\
400 & 4.3 & 0.1 & 7812.5 & 29890.6 & 0.13 & 3583.3 & 29765.1 & 0.032\\
460 & 4.9 & 0.1 & 7812.5 & 63335.8 & 0.13 & 2693.4 & 61181.1 & $-0.024$
\\\hline\hline
 \end{tabular}
 \caption{Comparison of tree-level obtained parameters ($^0$) versus the ones 
computed at one loop for different values of the three inputs.}\label{tab:comparison}
 \end{center}
\end{table}
\begin{figure}[t]
 \includegraphics[width=\columnwidth]{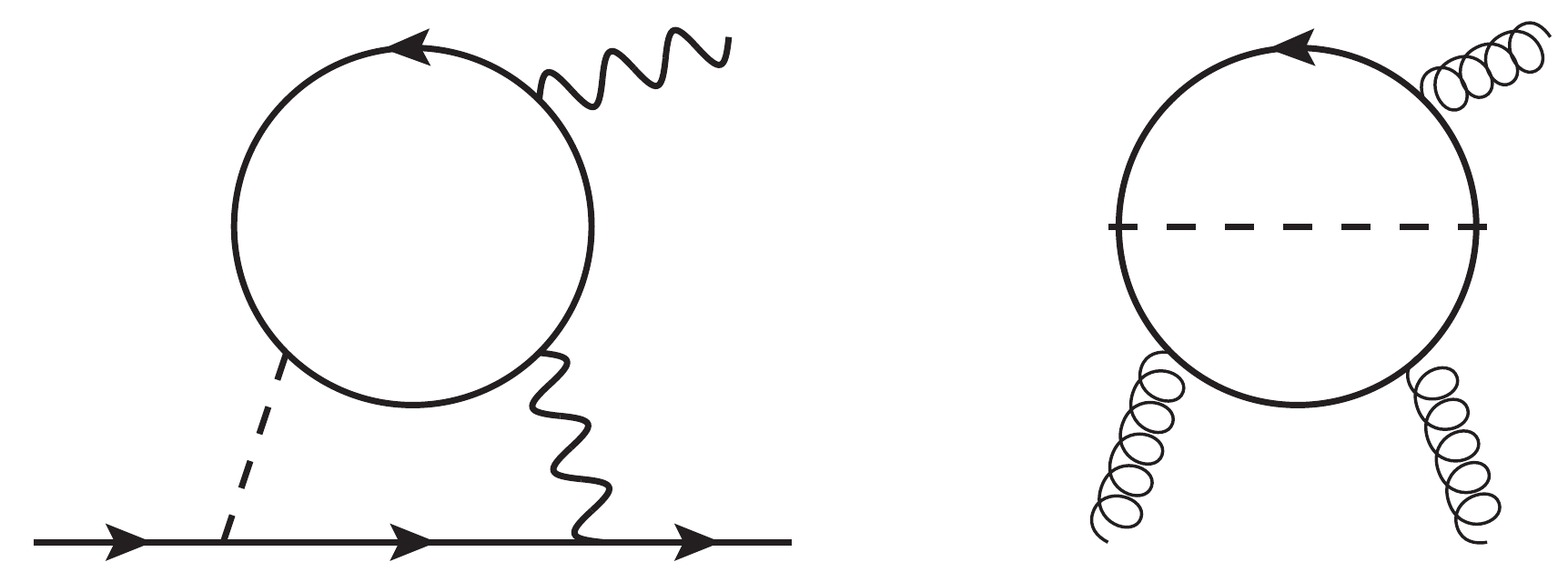}
 \caption{\it Main contribution to the electron (left) and neutron (right) EDMs in the scalar triplet 
extension of the SM with a low cutoff.}\label{fig:edms}
\end{figure}
\begin{figure*}[ht!]
 \begin{center}
  \includegraphics[width=1.8\columnwidth]{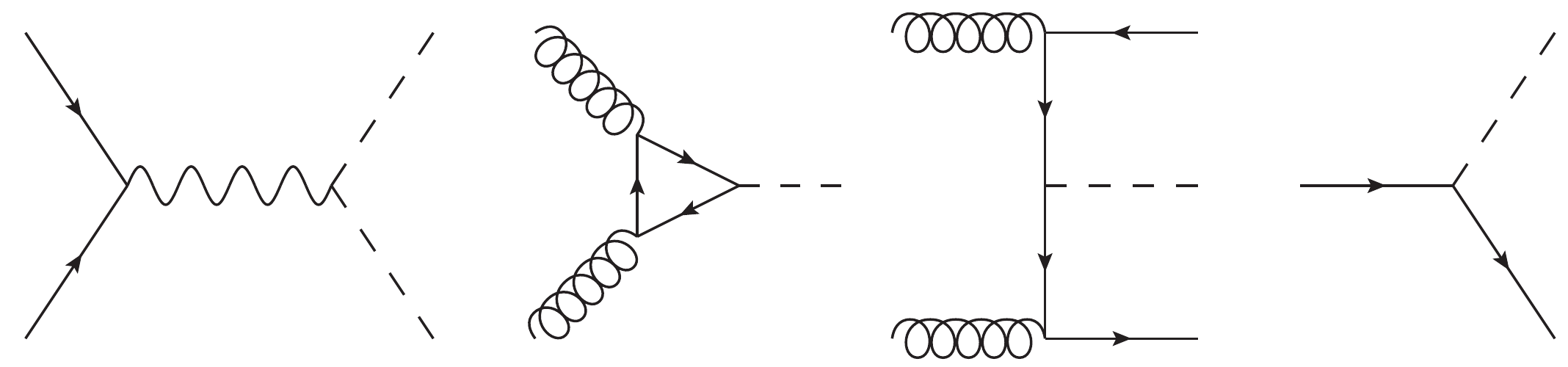}
 \end{center}
\caption{\it Representative diagrams of the main $\Phi$ production mechanisms 
at $pp$ colliders. Left) Pair production via EW currents. Middle-left) Single 
production via gluon fusion. 
Middle-right) Production in association with $t\overline{b}$ 
($\overline{t}b$). Right) Production from the decay of a top 
quark.}\label{fig:mechanisms}
\end{figure*}

Large values of $\lambda_{H\Phi}$ can be also probed 
in the 
$h\rightarrow\gamma\gamma$ channel. Indeed, the width of the later in this case 
reads
\begin{align}
 \Gamma(h\rightarrow\gamma\gamma) = \frac{\alpha^2 m_h^3}{1024 
\pi^3}\bigg\lbrace&\frac{2}{v}\bigg[A_1(\tau_W) + 
\frac{4}{3}A_{1/2}(\tau_t)\bigg] 
\\\nonumber
&+\lambda_{H\Phi}\frac{v}{m_\Phi^2}A_0(\tau_\Phi)\bigg\rbrace^2~,
\end{align}
with $\alpha$ the electromagnetic constant and $\tau_i = 
4m_i^2/m_h^2$. The last ATLAS+CMS combined measurement of the Higgs decay into 
photons was provided in Ref.~\cite{Sirunyan:2018koj}, 
$\Gamma(h\rightarrow\gamma\gamma)/\Gamma(h\rightarrow\gamma\gamma)_{
\text{SM}} = 1.14^{+0.19}_{-0.18}$. The region in 
the plane $(m_\Phi, \lambda_{H\Phi})$ that is consequently excluded at the $95$ 
\% CL is enclosed by the solid orange line in Fig.~\ref{fig:ewpt}. The 
expectation at the HL-LHC is that ratios outside the range $1.0\pm 0.1$ will be 
excluded~\cite{ATL-PHYS-PUB-2013-014}. The corresponding region is enclosed by 
the dashed orange line. It is clear that, if departures from the SM prediction 
on the Higgs to diphoton rate are not observed, only one-step EWPT 
(\textit{i.e.} single peak signatures) could be detected by LISA.

Finally, let us very briefly comment on the possibility of EW 
baryogenesis~\cite{Kuzmin:1985mm,Cohen:1993nk,Riotto:1999yt,Morrissey:2012db}.
In our scenario,  CP is violated 
spontaneously during the second transition in the two-step case,
when both $h$ and $\phi_0$ change VEV and therefore the top mass acquires a CP 
violating phase. 
In related models~\cite{Espinosa:2011eu} (see also 
Ref.~\cite{Vaskonen:2016yiu,Huang:2018aja,Grzadkowski:2018nbc}), EW baryogenesis has been 
shown 
successful provided 
$c\Delta v/f \gtrsim 0.1$, with $\Delta v$ 
the change in VEV during the EWPT.
In our case, $\Delta v$ can be easily 
$\gtrsim 100$ GeV (see Fig.~\ref{fig:evol}) and therefore 
$c \Delta v/f \gtrsim 0.1$ for $c/f\sim 1$ TeV$^{-1}$. 

A small explicit CP violating potential $\Delta V/T_n^4 \gg \text{H}/T_n\sim 
10^{-16} $, with the Hubble scale $\text{H}$, is only 
needed to avoid domain wall problems~\cite{Espinosa:2011eu}.
In our setup, this can be triggered by a small CP-violating term in the 
potential, $\sim \kappa\Phi H^2$. At leading order in $\kappa$, it reflects in 
the (finite-temperature)
potential as $V\sim\kappa T^3/(4\pi)^2$. Avoiding domain 
walls then implies
$\kappa \gtrsim 10^{-12}$ GeV.

Let us show that this amount of CP violation evades easily neutron and electron 
dipole moment 
(EDM) constraints. Indeed, the neutron EDM arises mainly from the running of the diagram 
on the right panel of Fig.~\ref{fig:edms}. It 
gives~\cite{Choi:2016hro}
\begin{equation}
 \bigg|\frac{d_n}{e}\bigg|\sim 20 ~\text{MeV} \frac{g_3^3}{(4\pi)^4}\frac{y_t^2 
v^2}{m_t^2 m_\Phi^2} \frac{c\kappa}{f} h(m_t^2/m_h^2)~,
\end{equation}
with $g_3$ the QCD coupling at the scale $\sim 1$ GeV, and
\begin{align}
 h(z) &=\\\nonumber
 &=z^2 \int_0^1 dx \int_0^1 dy \frac{x^3 y^3 (1-x)}{[zx(1-xy) + 
(1-x)(1-y)]^2}~.
\end{align}
For the values of $c/f$ and $\kappa$ stated previously we 
obtain $|d_n|\sim 10^{-38}$ e\,cm, much smaller than the current 90\,\% CL 
bound 
$2.9\times \sim 10^{-26}$ e\,cm~\cite{Baker:2006ts}.

An electron EDM
will be generated 
mainly via two-loop diagrams as that depicted in the left panel 
of Fig.~\ref{fig:edms}.
Using the expressions of Ref.~\cite{Keus:2017ioh}, we 
find that 
the value of the electron EDM in our case reads
\begin{equation}
 \frac{d}{e} \sim \frac{\alpha v^2 c}{6\pi^3 m_t f} y_t \frac{\kappa 
y_e}{m_\Phi^2} 
\bigg[f(m_t^2/m_h^2) + g(m_t^2/m_h^2)\bigg]~, 
\end{equation}
with
\begin{align}
 f(z) &= \frac{1}{2}z\int_0^1 \frac{1-2x(1-x)}{x(1-x)-z}\log{\frac{x(1-x)}{z}} 
dx~,\\
 g(z) &=\frac{1}{2}z\int_0^1 \frac{1}{x(1-x) -z}\log{\frac{x(1-x)}{z}}dx~.
\end{align}
We obtain $|d_e|\sim 10^{-42}$ e\,cm, much smaller than
the latest measurement by ACME~\cite{Andreev:2018ayy}, $d < 1.1\times 
10^{-29}$ e\,cm. 

Regarding $c/f$, more stringent bounds could be set at colliders. We dedicate 
next section to this point.

\section{Collider signatures}\label{sec:collider}
\begin{figure*}[t]
 \begin{center}
  \includegraphics[width=\columnwidth]{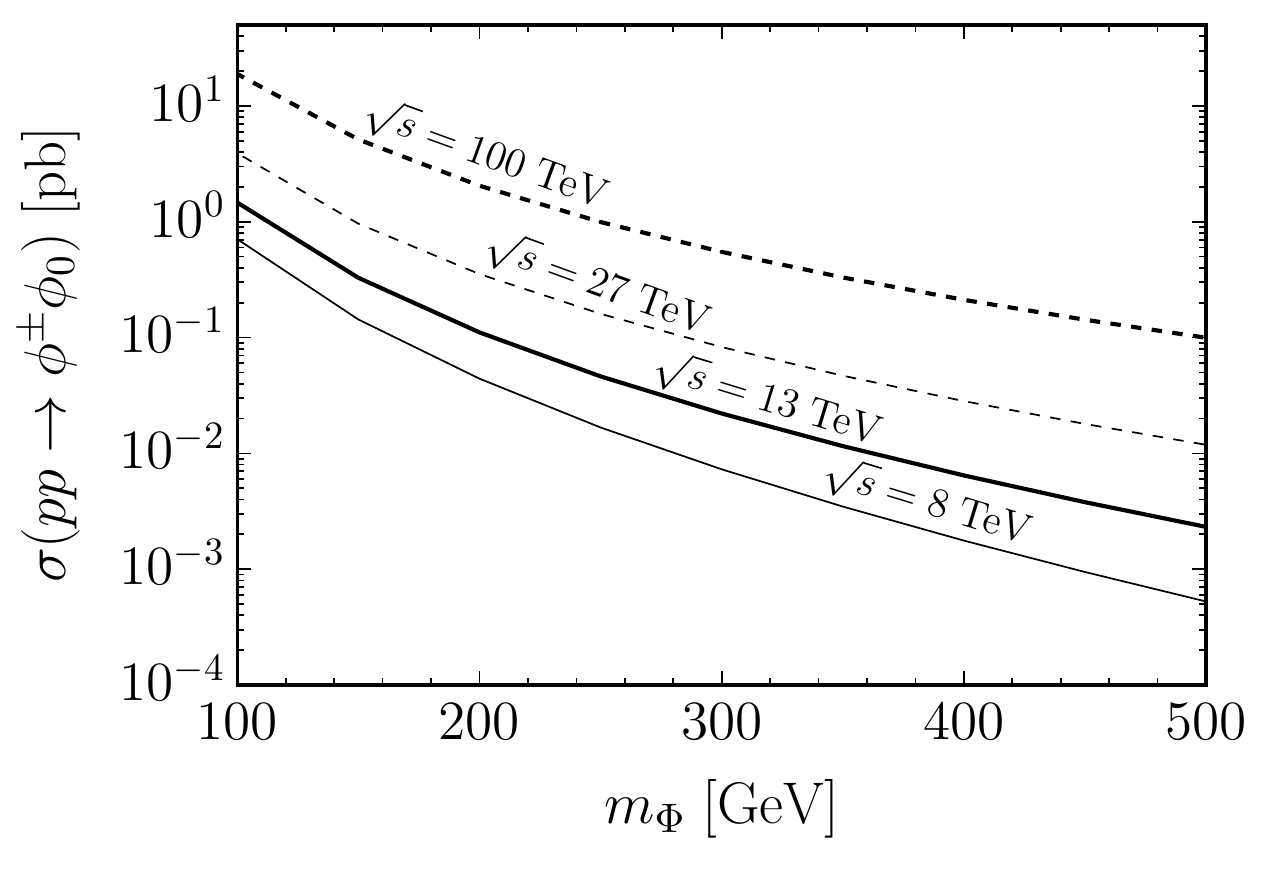}
  \includegraphics[width=\columnwidth]{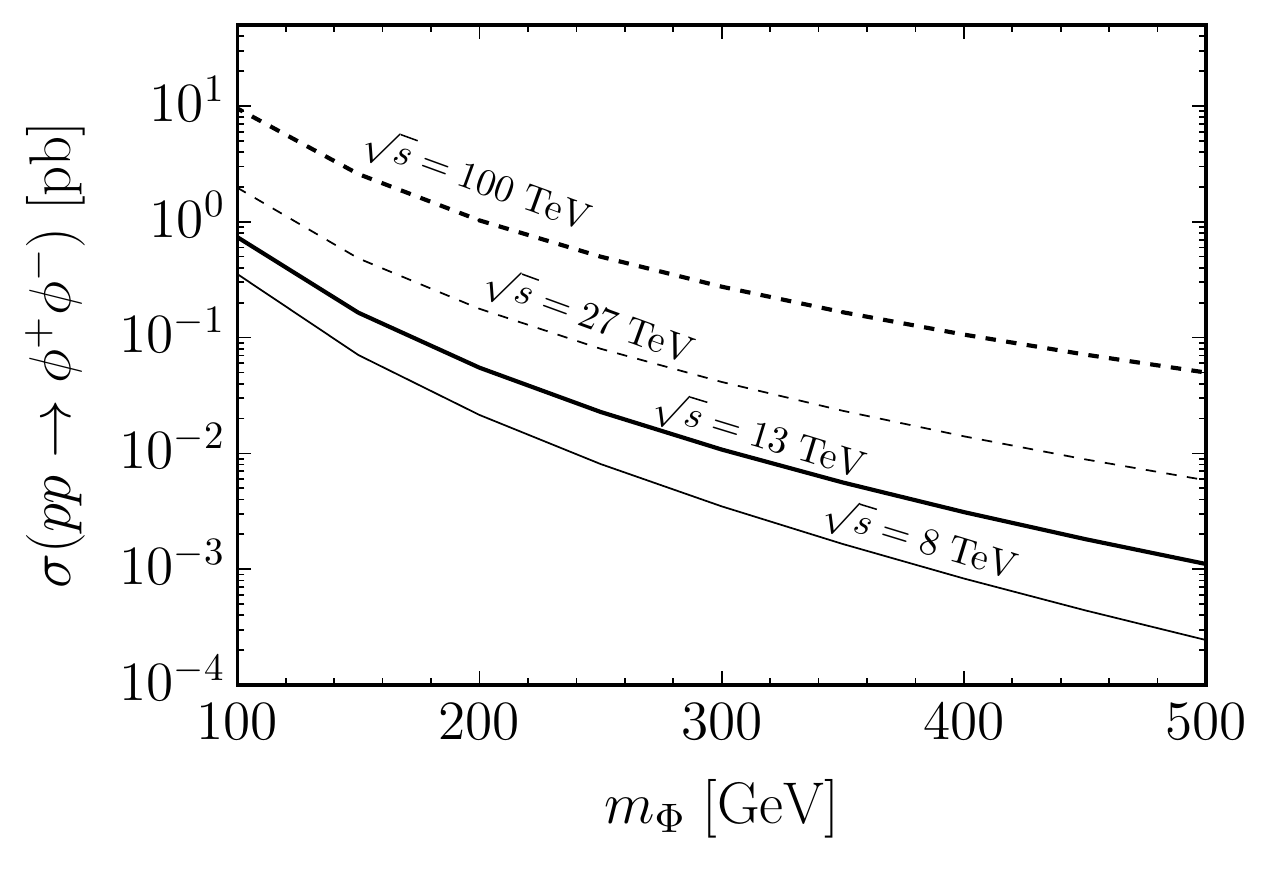}
  \includegraphics[width=\columnwidth]{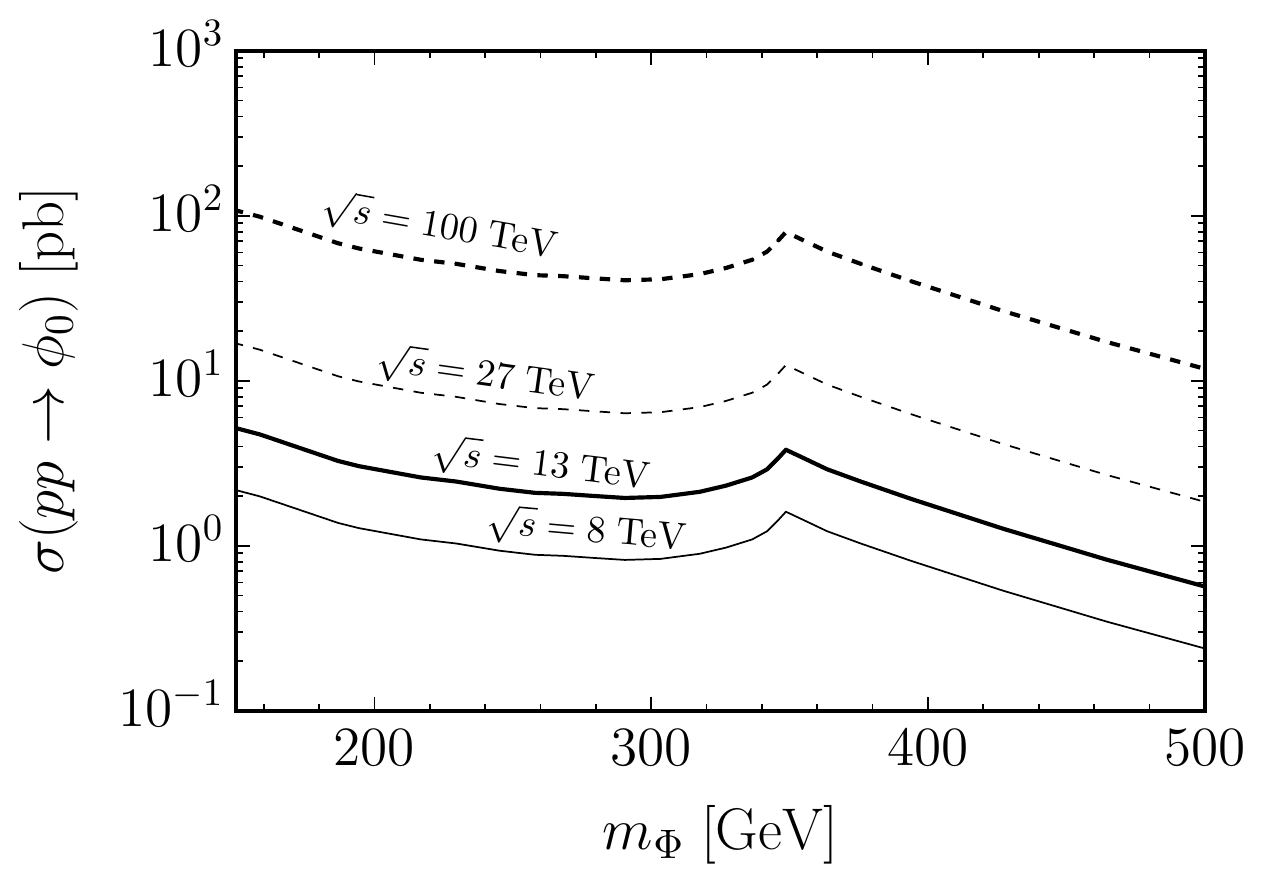}
  \includegraphics[width=\columnwidth]{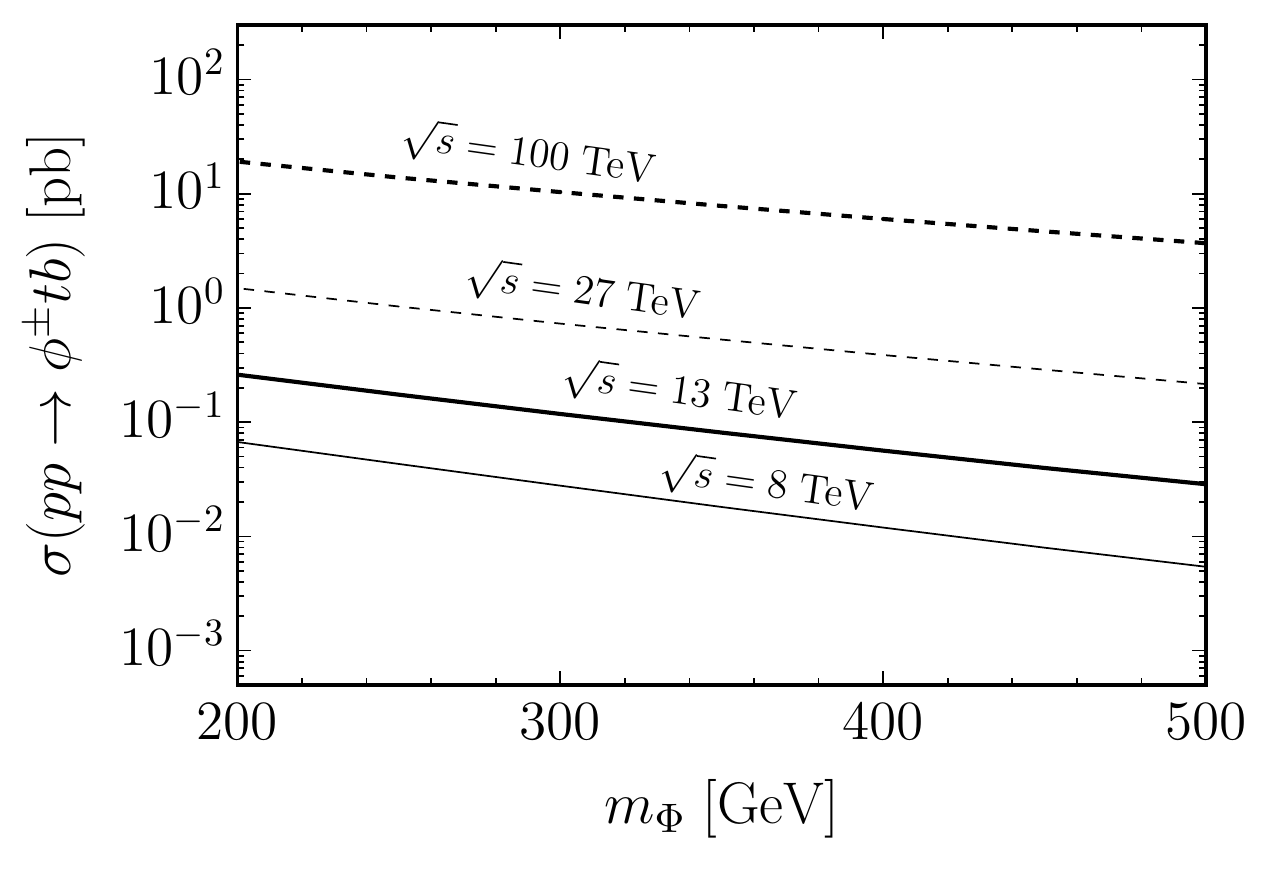}
 \end{center}
\caption{\it Cross section of the different production modes for $\Phi$ at $pp$ 
colliders of different c.m.e. The coupling $c/f$ is set to 
$1\rm{TeV}^{-1}$. The production cross section of $\phi_0$ via gluon fusion was 
rescaled from Ref. \cite{Primulando:2016eod}, at $\sqrt{s}=13$ TeV. To 
obtain the cross sections for other c.m.e., 
we have computed the corresponding ratio in \texttt{MadGraph}, by using the 
Higgs EFT of the \texttt{ggHFullLoop} model. This ratio turns out to be a good 
approximation of the ratio in the full theory; see Ref. 
\cite{Li:2010fu}.}\label{fig:xsecs}
\end{figure*}
The scalar triplet can be produced at $pp$ colliders in a variety of ways; see 
Fig.~\ref{fig:mechanisms}. The corresponding cross sections at $\sqrt{s} = 8, 
13$ 
TeV are given in Fig.~\ref{fig:xsecs}. For completeness, we also provide 
numbers for $27$ and $100$ TeV center-of-mass energy.

\begin{figure}[t]
 \begin{center}
  \includegraphics[width=\columnwidth]{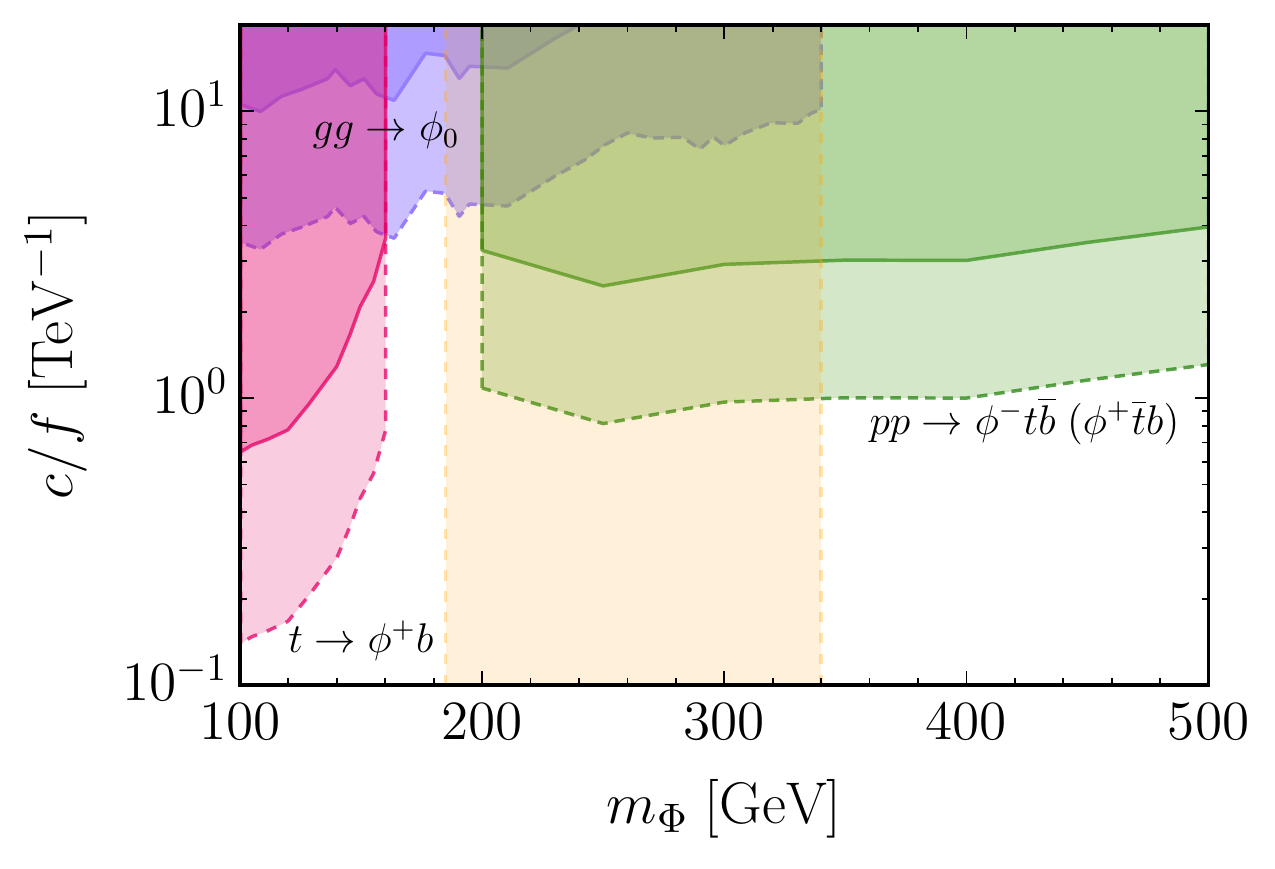}
 \end{center}
\caption{\it Current (solid) and future (dashed) bounds on the 
parameter space $(m_\Phi, c/f)$. The red region to the left comes 
from LHC searches for rare top decays, $t\rightarrow b\phi^\pm, 
\phi^\pm\rightarrow$ jets; see Ref.~\cite{Khachatryan:2015uua}. The upper 
blue region comes 
from LHC searches for single production of 
$\phi^0\rightarrow b\overline{b}$ via gluon fusion; see Ref.~\cite{Sirunyan:2018ikr}. 
The green region to the right comes from LHC searches for single production 
of $\phi^\pm$ in association with a top and a bottom 
quark; see Ref.~\cite{Aaboud:2018cwk}. The 
middle vertical orange band is the region that could be 
tested using our dedicated analysis.}\label{fig:colliders}
\end{figure}

The triplet can be singly produced in $q\overline{q}$ initiated 
processes. The Yukawa suppression, together with the $1/f$ factor, makes 
the production cross section in this 
channel very small, though. 
Still, $\phi_0$ 
can be singly produced in gluon fusion.
In the regime $m_\Phi < 2 m_t$, the most constraining 
searches are those looking for single production of $b\overline{b}$ resonances. 
The most up-to-date such analysis was recently released by CMS; see 
Ref.~\cite{Sirunyan:2018ikr}. It is based on $35.9$ fb$^{-1}$ of integrated 
luminosity collected at $\sqrt{s} = 13$ TeV. The region of the plane $(m_\Phi, 
c/f)$ that is excluded by this analysis is enclosed by the solid blue line 
in 
Fig.~\ref{fig:colliders}. It is expected that they become a factor of 
$\sqrt{3\,\text{fb}/35.9\,\text{ab}}\sim 9$ stronger at the HL-LHC. The 
projected bound on the plane is enclosed by the dashed blue line in the 
same 
figure. For $m_\Phi > 2m_t$, $\phi_0$ 
decays mostly 
into $t\overline{t}$. There are however no resonant searches for invariant 
masses below $500$ GeV, neither at $\sqrt{s}=8$ TeV nor $\sqrt{s}=13$ TeV. 

Moreover, the scalar triplet can be produced in association with top and bottom 
quarks, namely $pp\rightarrow \phi^{+} t \overline{b}$. For $m_\Phi > m_t$, the 
most updated and constraining search
is the ATLAS study of Ref.~\cite{Aaboud:2018cwk}, which uses $36.1$ 
fb$^{-1}$ of LHC data collected 
at 13 TeV. It combines both the semi- and di-leptonic channels. The limits on
$\sigma(pp\rightarrow tb\phi^\pm)\times\mathcal{B}(\phi^\pm\rightarrow tb)$ 
translate into the bounded region delimited by the green
solid line in Fig.~\ref{fig:colliders}.
A naive rescaling with the luminosity enhancement suggests that cross sections a 
factor of
$\sim 0.1$ smaller can be tested at
the HL-LHC. Translated to the plane $(m_\Phi, c/f)$, the corresponding bound is 
given
by the region enclosed by the dashed line of the same colour.

In addition, for $m_t > m_\Phi$, the triplet can be also produced in the decay 
of 
the top quark. Current searches for $t\overline{t}$ production with 
$t\rightarrow \phi^\pm b, \phi^\pm\rightarrow jj$ have been carried out in CMS 
at $\sqrt{s}=8$ TeV with an integrated luminosity of $\mathcal{L} = 19.7$ 
fb$^{-1}$~\cite{Khachatryan:2015uua}. This latter reference sets an upper bound 
on this rare top decay of $\mathcal{B}(t\rightarrow \phi^\pm b, 
\phi^\pm\rightarrow j j) <$ 1--2 \% for $m_\Phi \sim$ 100--160 GeV. Using 
Eq.~\ref{eq:widths}, this constraint translates into the region enclosed by the 
solid 
red line in Fig.~\ref{fig:colliders}. The projected bound at the HL-LHC is 
depicted, too.

Finally, irrespectively of the value of $c/f$, the scalar triplets can be 
always pair-produced via EW charged currents (CC), 
$pp\rightarrow W^{\pm(*)}\rightarrow\phi^\pm\phi_0$, as well as via neutral 
currents (NC), $pp\rightarrow Z/\gamma\rightarrow\phi^+\phi^-$.  (Note that 
$\phi_0$ 
does not interact with the $Z$ boson and therefore it can not be 
pair-produced via NCs.) For $m_\Phi < m_t$, the new charged
and neutral scalars decay mainly into $q'\overline{q}$ and $b\overline{b}$,
respectively. Searches for pair-produced dijet resonances might therefore be
sensitive to this regime. The most constraining such search is the CMS analysis
presented in Ref.~\cite{Sirunyan:2018rlj}. At this mass scale, each pair of 
quarks is
very collimated and manifests as a single jet. The experimental analysis
uses boosted techniques, including jet grooming to remove
QCD radiation.
The analysis considers $\mathcal{L} = 35.9$ fb$^{-1}$ at 13 TeV of c.m.e. The 
current limits on the total cross section range from $\sim 170$ pb (100 GeV) 
to $\sim 20$ pb (170 GeV). Therefore, 
the parameter space of our model is not constrained. Furthermore, a naive 
rescaling with the larger luminosity shows that this analysis will not be even 
constraining at the HL-LHC. 

For 
$m_t<m_\Phi<2m_t$, the NC process gives rise to the final state $t\overline{b}, 
\overline{t}b$. The latest analysis exploring this channel for masses below 
$500$ GeV was performed by CMS at $\sqrt{s} = 8$ TeV; see 
Ref.~\cite{CMS:2013gqa}. Unfortunately, the corresponding limits range 
from $\sim 2.5$ pb (250 GeV) to $\sim 0.5$ pb (500 GeV). No 
region in our parameter space can be even 
constrained at the HL-LHC.
Likewise, the CC gives $t\overline{b}(\overline{t}b), 
b\overline{b}$. To the best
of our knowledge, there is however no dedicated 
search for pair produced resonances decaying to these final states.
Being this channel $c/f$ independent, we perform a signal and background 
simulation of this process in section~\ref{sec:analysis}.

For 
$m_\Phi > 2m_t$, the NCs still give resonant $t\overline{b}, b\overline{t}$. 
The CC channel instead results in $t\overline{t}, t\overline{b} 
(b\overline{t})$. Once more, no dedicated analysis exists for this final state.
(The lack of analyses sensitivity to similar final states has been also 
recently pointed out in Ref.~\cite{Kribs:2018ilo} in the context of composite 
dark sectors.) However, in comparison to this one, the 
$t\overline{b}(b\overline{t}),b\overline{b}$ analysis is much cleaner. 
Furthermore, it probes the mass range 
where the 2-step EWPT, and therefore EW baryogenesis, can occur.

\section{LHC sensitivity}\label{sec:analysis}
EW pair production of $t\overline{b} (\overline{t}b), b\overline{b}$ 
resonances occurs naturally in 
broadly-studied models, such as the 2HDM. Moreover, the cross section is 
independent of scalar to fermion couplings, provided the former decay promptly. 
It is therefore surprising that no experimental search has explored this 
channel at the LHC yet.

A plausible explanation is that the majority of these 
analyses are based on the 2HDM of the MSSM. In that case, 
one Higgs doublet gives mass to the up fermions, while the second gives mass to 
the down fermions. The physical charged components then couple to the top and 
bottom quarks with effective $c/f \sim (y_t\cot{\beta} + y_b\tan{\beta})/v$, 
with $\tan{\beta}$ the ratio of the two doublet VEVs. Therefore, $c/f \geqslant
2\sqrt{y_b y_t}/v\sim 1$ TeV$^{-1}$, the inequality being saturated at 
$\tan{\beta} = 
\sqrt{y_t/y_b}\sim 1.3$.  As it can be seen in Fig.~\ref{fig:colliders}, this value
is at the reach of $b\overline{b}$ resonant searches. Therefore, there is \textit{a priori}
no need for further analyses. The situation in our model and in other versions of
the 2HDM is very different, though, because the coupling of the new scalars to 
the SM fermions can be (and typically is) very small.
In the composite Higgs scenario, the effective scalar-fermion coupling can also be small,
\textit{i.e.} $c/f \lesssim 1~\text{TeV}^{-1}$, according to equation \ref{Lchm} for natural values of 
$0 < \gamma \leq 1$.

With the aim of filling this gap, we perform a dedicated analysis for the 
process $pp\rightarrow \phi^\pm \phi_0\rightarrow t\overline{b} (\overline{t}b), 
b\overline{b}$.
We generate the hard processes for signal and background using 
\texttt{MadGraph v5}~\cite{Alwall:2014hca} and 
shower them to a fully hadronised final state using \texttt{Pythia 
v8}~\cite{Sjostrand:2014zea}.
No parton-level cuts are applied. The main backgrounds are 
$t\overline{t}+\text{jets}$, $t\overline{t} b\overline{b}$, $t(\overline{t})+3b$ 
and $W + 4b$. 

\begin{figure}[t]
 \begin{center}
  \includegraphics[width=\columnwidth]{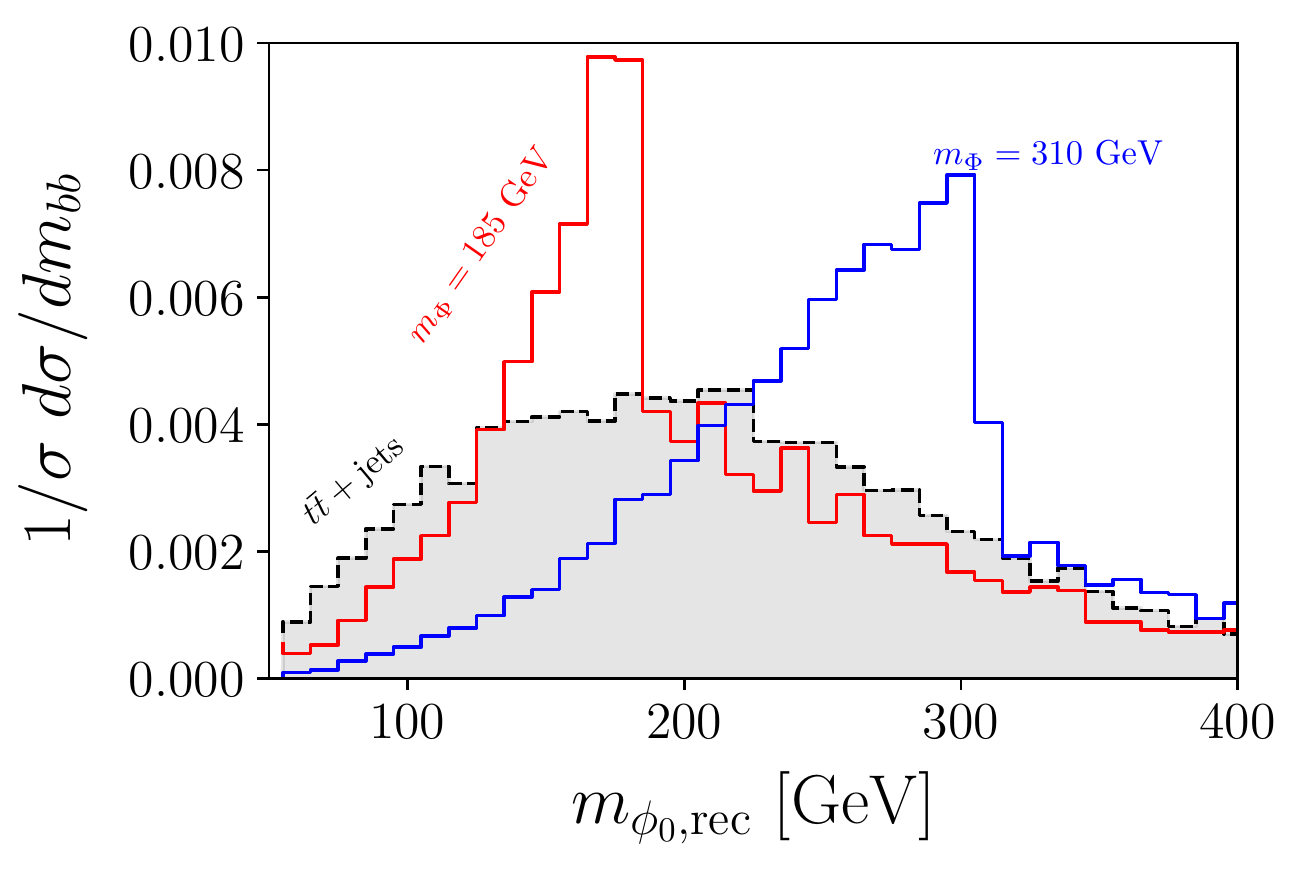}
 \end{center}
\caption{\it Normalized distribution of $m_{\phi_0,\mathrm{rec}}$ in the main 
background (dashed black) and in the signal for $m_\Phi = 185$ 
GeV (solid red) and $m_\Phi = 310$ GeV (solid blue).}\label{fig:mbb}
\end{figure}

\begin{figure}[t]
 \begin{center}
  \includegraphics[width=\columnwidth]{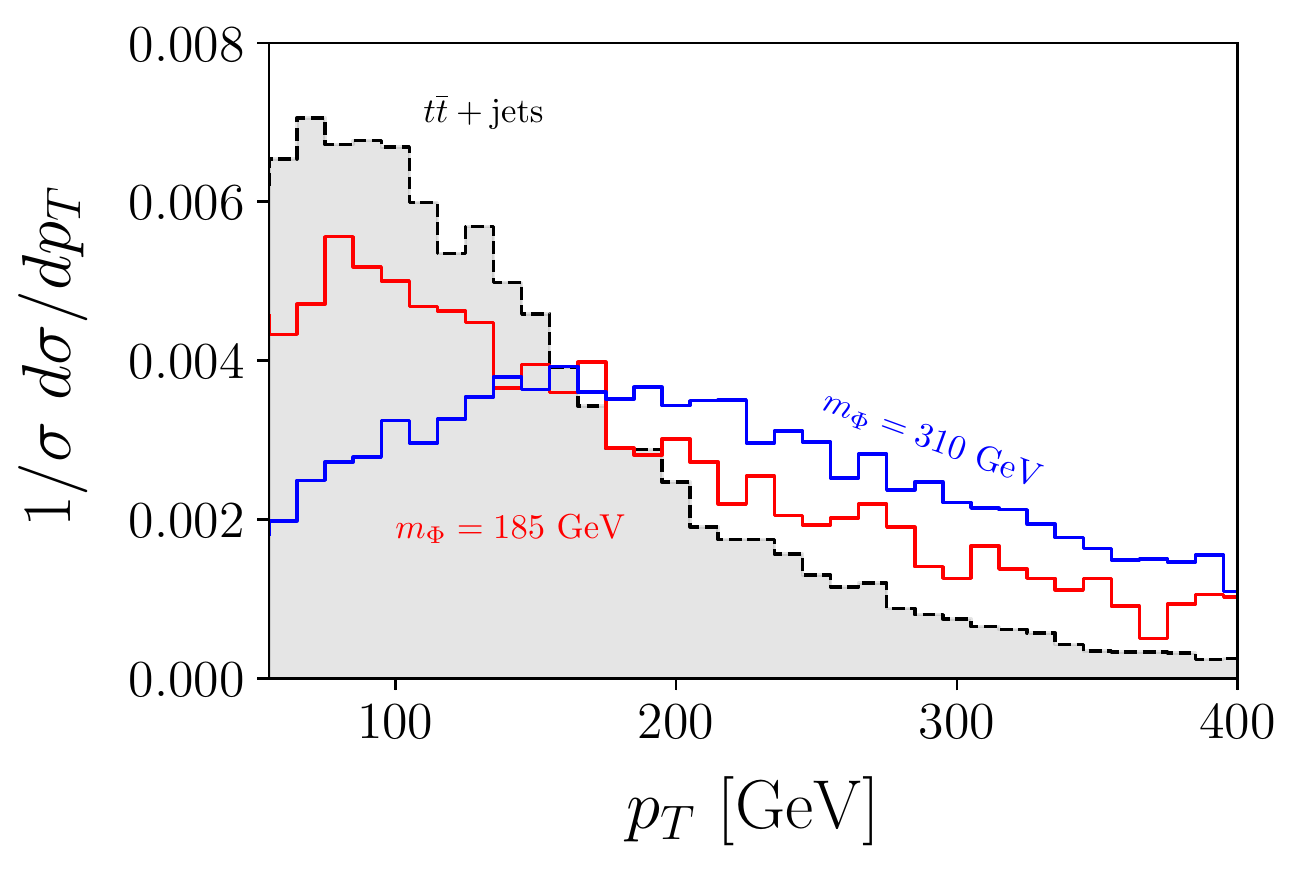}
 \end{center}
\caption{\it Same as Fig.~\ref{fig:mbb} but for $p_T$ of the 
reconstructed $\phi^0$.}\label{fig:pt}
\end{figure}

At the reconstruction level, a lepton is considered isolated if the hadronic 
energy deposit within a cone of size $R = 0.3$ is smaller than 10\,\% 
of the lepton candidate's $p_T$. Jets are 
defined by the anti-$k_t$
algorithm with $R=0.4$. The following cuts are imposed:
\begin{enumerate}
 \item Exactly one isolated lepton with $|y| < 2.5$ and $p_T > 15$ GeV;
 \item At least four jets, with $p_T > 30$ GeV.
\end{enumerate}
The longitudinal component of the missing neutrino is reconstructed using the 
requirement $m_W^2 = (p_l+p_\nu)^2$, with $m_W$ the mass of the $W$ boson. The 
neutrino and lepton four-momenta, $p_\nu$ and $p_l$, are then 
added to the jet wich gives a total invariant mass closest to the top quark 
mass. After this,
\begin{enumerate}
 \setcounter{enumi}{2}
 \item The invariant mass of this top is required to be within $50$ GeV of the 
top mass;
\item Three $b$-tagged jets are to be found among the jets not coming from the 
leptonic top.
\item We require the reconstructed masses of $\phi_0$ and $\phi^\pm$ to be 
similar,
\textit{i.e.} $|m_{\phi_0,\mathrm{rec}} - m_{\phi^\pm,\mathrm{rec}}| \leq 50$ 
GeV.
\end{enumerate}
To decide which $b$-tagged jet is assigned to $\phi^\pm$ and which two are assigned to 
$\phi_0$, we compute all possible combinations and choose the one resulting in the 
minimum difference between $m_{\phi_0,\mathrm{rec}}$ and $m_{\phi^\pm,\mathrm{rec}}$.
The normalized distribution of this former variable in the main background 
($t\overline{t}$+jets) and in the signal for $m_\Phi = 185$ GeV and 
$m_\Phi=310$ GeV is depicted in Fig.~\ref{fig:mbb}. In Fig.~\ref{fig:pt}, we 
also show the normalized distribution of the $p_T$ of the reconstructed 
$\phi^0$. However, cutting on this variable is costly in cross section, but
would allow to improve the ratio of signal over background further.

Finally, 
\begin{enumerate}
\setcounter{enumi}{5}
 \item We reconstruct the resonances $\phi_0$ and $\phi^{\pm}$ in the mass 
window of $\pm 30$ and $\pm 40$ GeV respectively. As shown in 
Fig.~\ref{fig:mbb} the experimental width of the resonances depends on 
their masses, thus, the central value of the mass window in the 
$(m_{\phi_0,\mathrm{rec}}, 
m_{\phi^\pm,\mathrm{rec}})$ plane has to be optimised for each $m_\Phi$ 
separately.
\end{enumerate}
The cut-flow for the signal and the relevant backgrounds is given in 
Tab.~\ref{tab:cutflow}. The sensitivity estimates are conservative, as the 
reconstruction relies on fairly inclusive cuts using large mass windows. 
Further, it is likely that some of the background was counted twice, as 
$b$-quarks from the parton shower in $t\bar{t}$ and from the matrix element in 
$t\bar{t} b\bar{b}$ are both contributing to the total background. Thus, we 
expect that the sensitivity can be improved further using a combination 
of multi-variate techniques \cite{Soper:2011cr, Soper:2012pb, Soper:2014rya} and 
high-$p_T$ final states \cite{Abdesselam:2010pt}.
\begin{figure*}[t]
 \includegraphics[width=0.68\columnwidth]{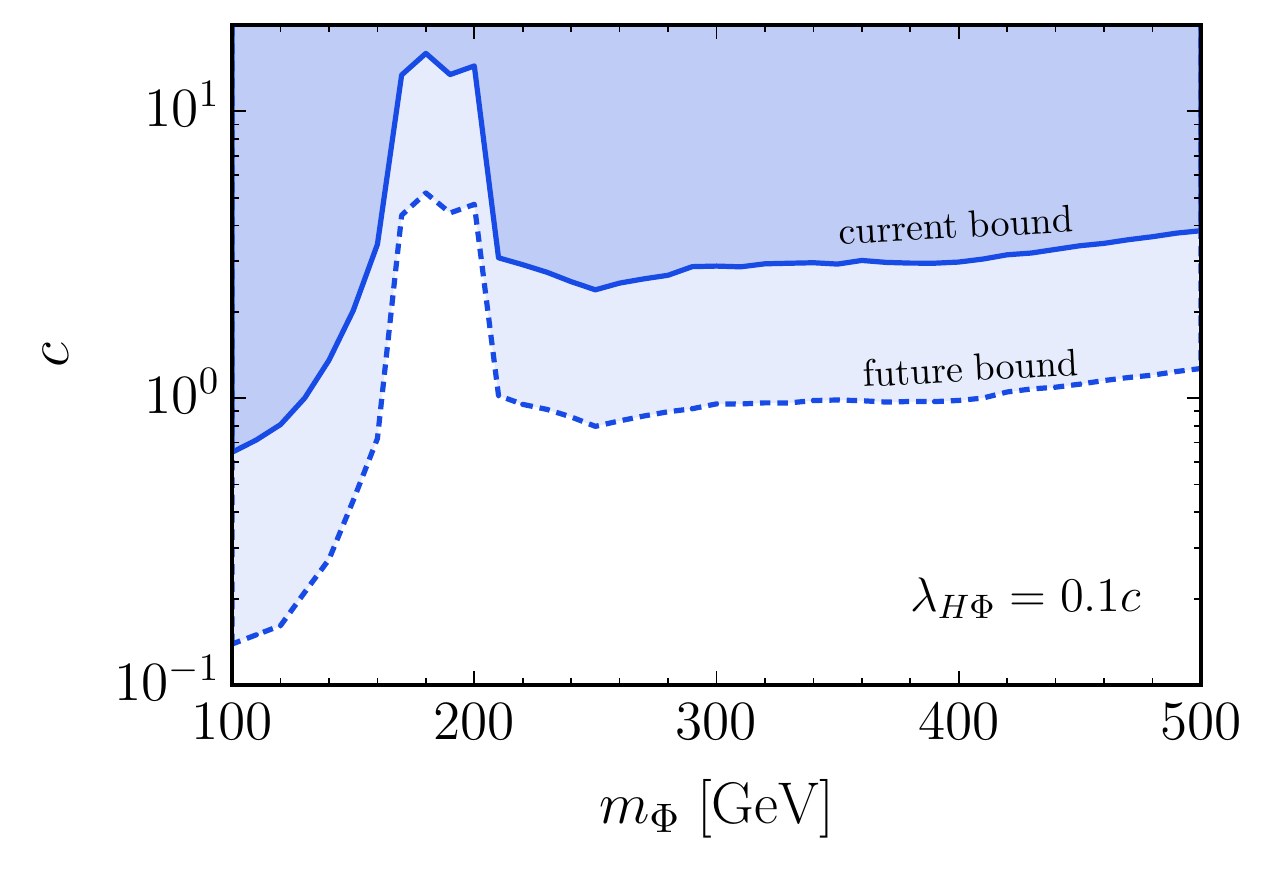}
  \includegraphics[width=0.68\columnwidth]{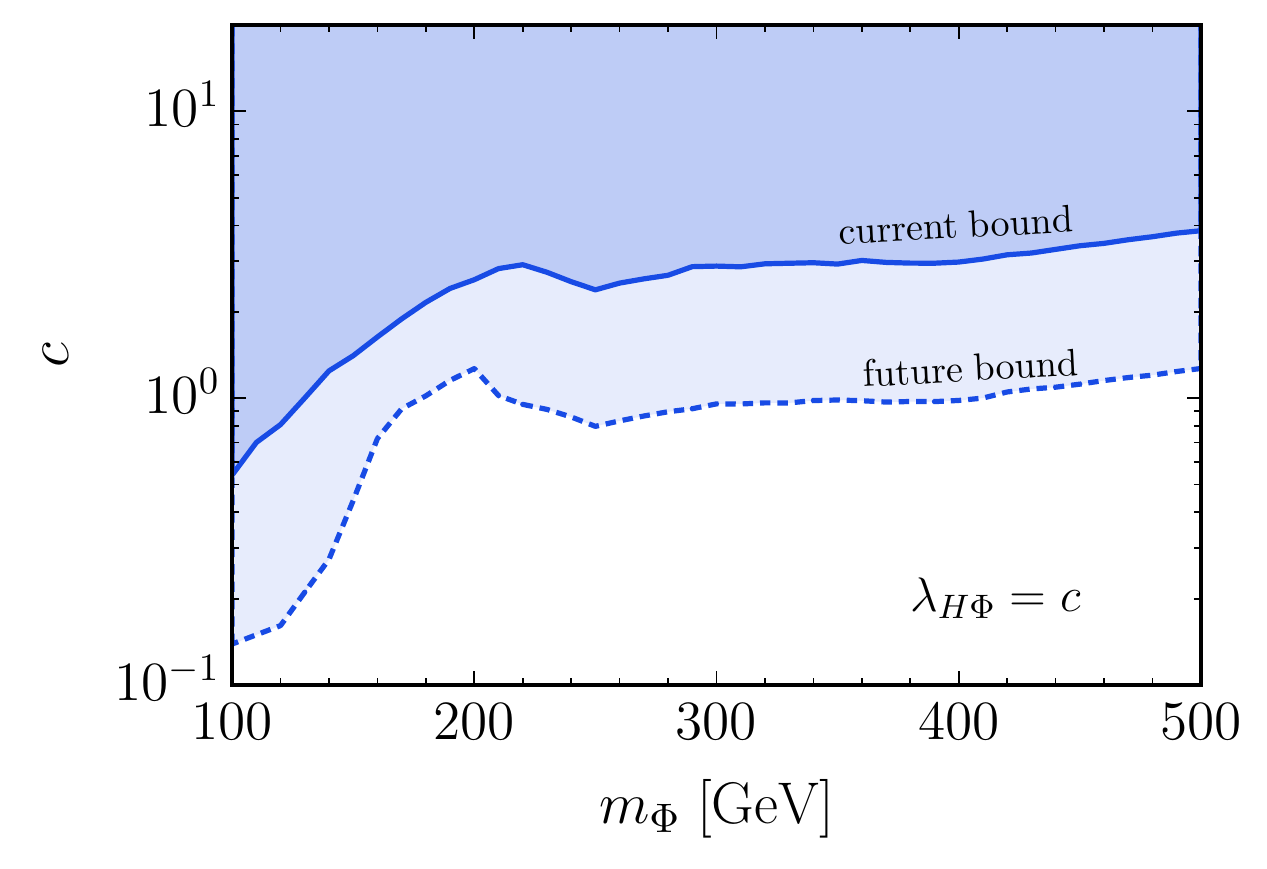}
 \includegraphics[width=0.68\columnwidth]{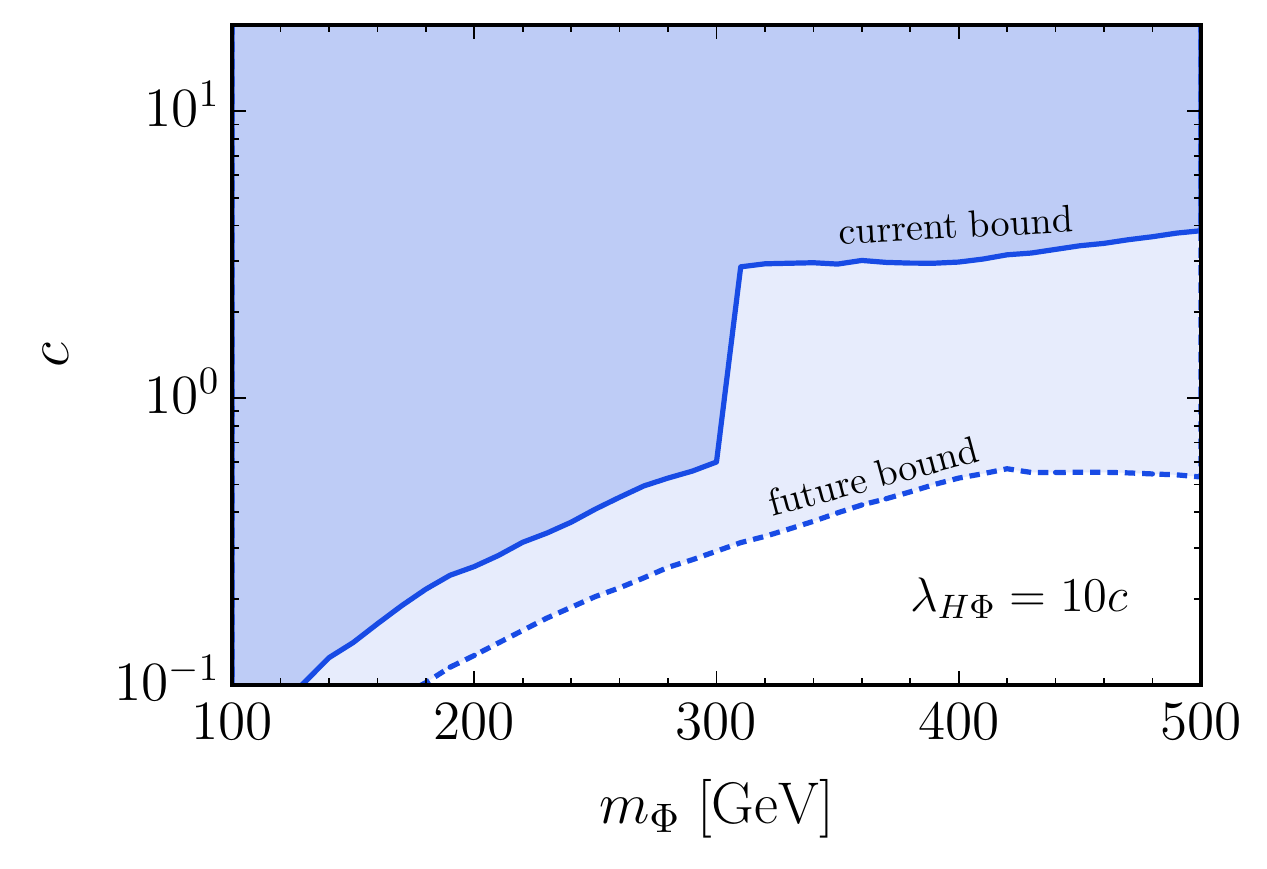}
 \caption{\it Current (solid) and futute (dashed) bounds on 
the plane $(m_\Phi, c)$ for $\lambda_{H\Phi} = 0.1c$ (left), 
$\lambda_{H\Phi} = c$ (center) and $\lambda_{H\Phi}=10c$ 
(right) and $f=1$ TeV.}\label{fig:final}
\end{figure*}
\begin{table}[t]
 \begin{tabular}{c|ccccc}
  \hline\hline
 Cuts & $m_\Phi = 185$ & $t\overline{t}+$ jets  & $t+3b$ & $t\overline{t} b\overline{b}$ &$W + 4b$    \\\hline
iso. lepton & 72.94 & 96693.1 & 0.65632 & 326.72& 1.327   \\
nr. jets  & 29.71 & 55288.5 & 0.6834 &  305.10 & 0.6147 \\
lep. top & 17.69 & 32626.6 & 0.393 & 198.10 & 0.3647   \\
$3~ b$-tags & 2.3 &267.4 & 0.07531& 45.45 & 0.0835  \\
similar mass & 0.93 & 81.1 & 0.0235 & 12.35 & 0.0220  \\
  \hline\hline
 \end{tabular}\vspace{0.5cm}
  \begin{tabular}{c|cc|c }
  \hline\hline
 final rec   &  signal &  background  & $s/\sqrt{s+b}$  \\\hline
$\Phi(185)$ & 0.39 & 25.6 & 4.2    \\
$\Phi(235)$ & 0.78 & 33.5& 7.3    \\
$\Phi(285)$ & 0.41 & 26.5 & 4.3    \\
$\Phi(335)$ & 0.22 & 19.0& 2.7    \\

  \hline\hline
 \end{tabular}
 
 \caption{\it Top) Effective cross section in fb for the 
signal (for $m_\Phi = 185$ GeV) and the backgrounds after 
each cut (1 -- 5), as described in the text. Bottom) Effective cross section 
in fb after all 
cuts, including cut 6, for different signals and for the total 
background. The sensitivity at the HL-LHC is also shown.}\label{tab:cutflow}
\end{table}
We estimate the sensitivity at the HL-LHC as $\mathcal{S} = s/\sqrt{s+b}$, with $s$ and $b$ the 
number of signal and background events after all cuts, respectively. It ranges 
from 2.7 to 7.3 for 
$m_\Phi$ between  $185$ and $340$ GeV. Thus, at the LHC 
($\sqrt{s}=13$ TeV) we can probe the entire mass interval with 
$3000~\mathrm{fb}^{-1}$. The corresponding region in the plane 
$(m_\Phi, c/f)$ is therefore a vertical band, the one enclosed by the dashed 
orange line in Fig.~\ref{fig:colliders}.

\section{Conclusions}\label{sec:conclusions}
Natural scalar extensions of the SM must have a low cutoff $f$ preventing large 
corrections to the scalar masses. However, such models are usually studied 
neglecting $1/f$ terms.  Basing on the real triplet 
extension of the SM, we have highlighted that, if these terms are taken into 
account, the phenomenology can be drastically different. In particular, the 
only renormalizable interaction allowing the new scalars to decay is so 
suppressed by the measurement of the $\rho$ parameter, that decays mediated by 
effective operators dominate.

We have studied the reach of current LHC analyses. We have found that, despite 
being $f$ independent, searches for EW pair-produced charged scalars decaying 
to third generation quarks are absent. This is particularly surprising given 
that such signals appear in a plethora of new physics models, including the 
2HDM/MSSM. Therefore, we have developed a dedicated analysis to probe the 
cleanest of these channels: $pp\rightarrow \phi^\pm\phi_0\rightarrow 
t\overline{b} (\overline{t} b), b\overline{b}$. We have shown that the whole 
range 
of masses $\sim 185$--$340$ GeV can be tested at the HL-LHC. 

For this analysis, we have neglected new scalar couplings 
to the leptons. Under the sort of Minimal Flavour 
Violation~\cite{DAmbrosio:2002vsn} assumed after Eq.~\ref{eq:lag}, explicitly 
reproduced in concrete models as shown in Eq.~\ref{Lchm}, the only relevant 
lepton would be the tau. Still, the decay of $\phi^0$ into $\tau^+\tau^-$ would 
involve only a $\sim m_\tau^2/(m_b^2 N_c)\sim 5\,\%$ of its width; our results 
being effectively unaffected. If couplings to the leptons are accidentally 
larger, the scalar could be better seen elsewhere; see 
Ref.~\cite{Akeroyd:2018axd}.

We also stress that, had he assume flavour-violating couplings 
$c_{ij}^{u(d)}$, they would give rise to a plethora of signals in meson 
decays~\cite{Harnik:2012pb}. They could be also seen 
in top decays as in the singlet extension of the Higgs 
sector~\cite{Banerjee:2018fsx}.

On another front, we have studied the reach of the future gravitational wave 
observatory LISA to the gravitational waves produced in the EWPT for certain 
region of the parameter space of the model. In particular, we have demonstrated 
that regions not yet excluded by Higgs to diphoton measurements will be 
testable. In this region, the EWPT proceeds 
mainly in one step, and therefore only one signal peak might be expected.

Finally, it is worth mentioning that in concrete models $c$ and 
$\lambda_{H\Phi}$ related; normally $\lambda_{H\Phi}\propto c$. This is evident 
for example in CHMs, in which the former (latter) is induced by
integrating out heavier resonances at tree level (one loop). To exhaust this 
point, we plot in Fig.~\ref{fig:final} the current and future bounds on the 
plane $(m_{\Phi}, c)$ considering all collider searches and gravitational wave 
signatures for different simple 
assumptions on the relation $\lambda_{H\Phi} = \lambda_{H\Phi}(c)$.

\section*{Acknowledgments}
\noindent
We would like to thank Nuno Castro, Marek Lewicki, Mariano 
Quir\'os and Carlos Tamarit for 
helpful discussions. 
MC 
is supported by the Royal Society under the Newton International Fellowship 
programme. MR is supported by Funda\c{c}\~ao para a Ci\^encia e Tecnologia (FCT)
under the grant PD/BD/142773/2018 and also acknowledges financing from LIP 
(FCT, COMPETE2020-Portugal2020, FEDER, POCI-01-0145-FEDER-007334).

\noindent

\appendix

\section{Loop functions}
In our case, $2m_\Phi > m_h$, and therefore
\begin{align}
 A_0(x) &= -x^2 \bigg[x^{-1} - f(x^{-1})\bigg]~,\\\nonumber
 A_{1/2}(x) &= 2 x^2 \bigg[x^{-1} + (x^{-1} - 1)f(x^{-1})\bigg]~,\\\nonumber
 A_1(x) &= -x^2 \bigg[ 2 x^{-2} + 3 x^{-1} + 3(2 x^{-1} -1)f(x^{-1}) 
\bigg]~\nonumber
\end{align}
with
\begin{equation}
 f(x) = \arcsin^2{\sqrt{x}}~.
\end{equation}

\bibliographystyle{style}
\bibliography{notes}{}

\end{document}